\documentclass[a4paper,fleqn,useAMS,usenatbib]{mnras}

\usepackage{newtxtext,newtxmath}
\usepackage[T1]{fontenc}
\usepackage{ae,aecompl}

\usepackage{amssymb}
\usepackage{aas_macros}
\usepackage{natbib}
\usepackage{times}
\usepackage{graphicx}
\usepackage{xspace}

\newcommand{\sub}[1]{_{\rm #1}}
\newcommand{\up}[1]{^{\rm #1}}

\newcommand{\HI}{\hbox{H{\sc i}}\xspace}
\newcommand{\SiIII}{\hbox{Si{\sc iii}}\xspace}
\newcommand{\SiIV}{\hbox{Si{\sc iv}}\xspace}
\newcommand{\CIII}{\hbox{C{\sc iii}}\xspace}
\newcommand{\CIV}{\hbox{C{\sc iv}}\xspace}
\newcommand{\OI}{\hbox{O{\sc i}}\xspace}

\newcommand{\OIII}{\hbox{O{\sc iii}}\xspace}

\newcommand{\OV}{\hbox{O{\sc v}}\xspace}
\newcommand{\OVI}{\hbox{O{\sc vi}}\xspace}
\newcommand{\OVII}{\hbox{O{\sc vii}}\xspace}
\newcommand{\OVIII}{\hbox{O{\sc viii}}\xspace}
\newcommand{\OIX}{\hbox{O{\sc ix}}\xspace}
\newcommand{\NeVIII}{\hbox{Ne{\sc viii}}\xspace}
\newcommand{\MgX}{\hbox{Mg{\sc x}}\xspace}

\newcommand{\mstar}[2]{$M\sub{\ast} = #1 \times 10^{#2} \, \mathrm{M}\sub{\odot}$}

\newcommand{\mstarsim}[1]{$M\sub{\ast} \sim 10^{#1} \, \mathrm{M}\sub{\odot}$}

\newcommand{\eddratio}{$L / L\sub{Edd}$\xspace}
\newcommand{\tagn}{$t\sub{AGN}$\xspace}
\newcommand{\fduty}{$f\sub{duty}$\xspace}
\newcommand{\eddratiom}{L / L\sub{Edd}}
\newcommand{\tagnm}{t\sub{AGN}}
\newcommand{\fdutym}{f\sub{duty}}

\newcommand{\N}[1]{$N\sub{#1}$}
\newcommand{\Nm}[1]{N\sub{#1}}
\newcommand{\deltaN}[1]{$\Delta \log\sub{10} \Nm{#1}$}
\newcommand{\deltaNm}[1]{\Delta \log\sub{10} \Nm{#1}}
\newcommand{\avN}[1]{$\left\langle \log\sub{10} \Nm{#1} \right\rangle\sub{t} (R)$}
\newcommand{\avNm}[1]{\left\langle \log\sub{10} \Nm{#1} \right\rangle\sub{t} (R)}
\newcommand{\avNfossil}[1]{$\left\langle \Delta \log\sub{10} \Nm{#1} \right\rangle\sub{t}\up{fossil} (R)$}
\newcommand{\avNfossilm}[1]{\left\langle \Delta \log\sub{10} \Nm{#1} \right\rangle\sub{t}\up{fossil} (R)}

\newcommand{\fcov}[1]{$f\sub{cov}\up{#1}$}
\newcommand{\fcovm}[1]{f\sub{cov}\up{#1}}
\newcommand{\avfcov}[1]{$\left\langle \fcovm{#1} \right\rangle\sub{t} (R)$}
\newcommand{\avfcovm}[1]{\left\langle \fcovm{#1} \right\rangle\sub{t} (R)}
\newcommand{\avfcovfossil}[1]{$\left\langle \Delta \fcovm{#1} \right\rangle\sub{t}\up{fossil} (R)$}
\newcommand{\avfcovfossilm}[1]{\left\langle \Delta \fcovm{#1} \right\rangle\sub{t}\up{fossil} (R)}

\newcommand{\D}{\mathrm{d}}

\defcitealias{schaye_2015}{S15}

\title[Non-equilibrium metals in circumgalactic medium]{Metals in the circumgalactic medium are out of ionization equilibrium due to fluctuating active galactic nuclei}

\author[M. C. Segers et al.]{Marijke C. Segers,$^{1}$\thanks{E-mail: segers@strw.leidenuniv.nl}
Benjamin D. Oppenheimer,$^{2}$
Joop Schaye$^{1}$\newauthor
and Alexander J. Richings$^{3}$\\
$^{1}$Leiden Observatory, Leiden University, PO Box 9513, NL-2300 RA Leiden, the Netherlands\\
$^{2}$CASA, Department of Astrophysical and Planetary Sciences, University of Colorado, 389 UCB, Boulder, CO 80309, USA\\
$^{3}$Center for Interdisciplinary Exploration and Research in Astrophysics (CIERA) and Department of Physics and Astronomy,\\
Northwestern University, 2145 Sheridan Road, Evanston, IL 60208, USA
}

\date{Accepted XXX. Received YYY; in original form ZZZ}

\pubyear{2017}

\begin{document}
\label{firstpage}
\pagerange{\pageref{firstpage}--\pageref{lastpage}}
\maketitle

\begin{abstract}
We study the effect of a fluctuating active galactic nucleus (AGN) on the abundance of circumgalactic \OVI in galaxies selected from the EAGLE simulations. We follow the time-variable \OVI abundance in post-processing around four galaxies -- two at $z = 0.1$ with stellar masses of \mstarsim{10} and \mstarsim{11}, and two at $z = 3$ with similar stellar masses -- out to impact parameters of twice their virial radii, implementing a fluctuating central source of ionizing radiation. Due to delayed recombination, the AGN leave significant `AGN proximity zone fossils' around all four galaxies, where \OVI and other metal ions are out of ionization equilibrium for several megayears after the AGN fade. The column density of \OVI is typically enhanced by $\approx 0.3 - 1.0$~dex at impact parameters within $0.3R\sub{vir}$, and by $\approx 0.06 - 0.2$~dex at $2R\sub{vir}$, thereby also enhancing the covering fraction of \OVI above a given column density threshold. The fossil effect tends to increase with increasing AGN luminosity, and towards shorter AGN lifetimes and larger AGN duty cycle fractions. In the limit of short AGN lifetimes, the effect converges to that of a continuous AGN with a luminosity of $(\fdutym / 100 \%)$ times the AGN luminosity. We also find significant fossil effects for other metal ions, where low-ionization state ions are decreased (\SiIV, \CIV at $z = 3$) and high-ionization state ions are increased (\CIV at $z = 0.1$, \NeVIII, \MgX). Using observationally motivated AGN parameters, we predict AGN proximity zone fossils to be ubiquitous around $M\sub{\ast} \sim 10^{10 - 11} \, \mathrm{M}\sub{\odot}$ galaxies, and to affect observations of metals in the circumgalactic medium at both low and high redshifts.
\end{abstract}



\section{Introduction}
\label{sec:introduction}

Active galactic nuclei (AGN) play an important role in the formation and evolution of galaxies. Powered by the accretion of gas onto the central black hole \citep[BH; e.g.][]{salpeter_1964,lynden-bell_1969}, AGN are the most luminous objects in the Universe, releasing vast amounts of energy into the interstellar medium of their host galaxies and beyond. The various scaling relations between the properties of AGN and those of their host galaxies (see e.g. \citealt{kormendy+ho_2013} for a review), as well as the apparent tendency of AGN to reside in star-forming galaxies \citep[e.g.][]{lutz_2008,santini_2012}, suggest a close correlation between AGN activity and the star formation (SF) activity of the host. This is also supported by the remarkably similar evolution of the cosmic SF rate density and the cosmic BH accretion rate density \citep[e.g.][]{boyle+terlevich_1998,silverman_2008,mullaney_2012b}, which are both found to peak at $z \approx 2$. A correlation between AGN and SF activity is consistent with the prediction that both phenomena are fuelled by a common supply of cold gas \citep[e.g.][]{hopkins+quataert_2010}, as well as with observational evidence that AGN affect the SF in the host by acting as a local triggering mechanism \citep[e.g.][]{begelman+cioffi_1989,elbaz_2009}, and by regulating SF galaxy-wide \citep[e.g.][]{dimatteo_2005} as they drive galactic outflows (i.e. ejective feedback) and heat the gas in the halo (i.e. preventative feedback).

Furthermore, as powerful sources of radiation, AGN not only provide radiative feedback in the form of pressure and photoheating, they also affect the ionization state of the gas in and around the host galaxies. In particular, the abundance of neutral hydrogen (HI), as measured from the Ly-$\alpha$ absorption along the light-of-sight towards a quasar\footnote{Throughout this work, we will use the words `AGN' and `quasar' interchangeably.}, is observed to be suppressed in proximity to the quasar \citep[e.g.][]{carswell_1982,scott_2000,dallaglio_2008}, consistent with the expected local enhancement of the HI ionizing radiation field relative to the extra-galactic background. This effect is referred to as the line-of-sight proximity effect.

Using pairs of quasars, it is possible to probe the ion abundances in the circumgalactic medium (CGM) of a foreground quasar host in the transverse direction, by analysing the absorption in the spectrum of the background quasar. As studies of this transverse proximity effect generally find no reduction \citep[and in some cases even an enhancement; see e.g.][]{prochaska_2013} of the HI optical depth close to the foreground quasar \citep[e.g.][]{schirber_2004,kirkman+tytler_2008}, but do find effects of enhanced photoionization on the abundances of metal ions \citep[e.g. \CIV and \OVI; see][]{goncalves_2008}, it is clear that transverse proximity effects are not straightforward to interpret. Quasar radiation being anisotropic \citep[e.g.][]{liske+williger_2001,prochaska_2013} or the fact that quasars tend to live in overdense regions of the Universe \citep[e.g.][]{rollinde_2005,guimaraes_2007} might play a role. Nevertheless, these studies indicate that the response of HI to a local enhancement of the ionizing radiation field is vastly different from that of metal ions. While hydrogen has only two ionization states, such that the HI fraction decreases with an increasing ionization field strength, heavy elements like oxygen have multiple ionization levels, where the ion fractions peak at particular temperatures and densities that depend on the local photoionization rate.

The differences between HI and metal ions become even more evident when considering their behaviour in a fluctuating ionizing radiation field \citep{oppenheimer+schaye_2013b}. After the local radiation source has faded, the timescale on which ion species return to ionization equilibrium depends on the recombination timescale, as well as on the ion fraction of the recombined species in equilibrium: the latter can be close to one for metal ions, while being typically $\lesssim 10^{-4}$ for HI in the CGM. This leads to significantly longer `effective' recombination timescales for metals than for hydrogen, which are even further extended due to the multiple ionization levels that metals need to recombine through. \citet{oppenheimer+schaye_2013b} showed that, in contrast to \HI, metal ions at typical CGM densities can remain out of ionization equilibrium up to a few tens of megayears, as a result of delayed recombination after the enhanced AGN radiation field turns off. They define these out-of-equilibrium regions as AGN proximity zone fossils.

Both observations and theory indicate that the radiation output from AGN is not continuous, but rather happens in intermittent bursts. This is potentially due to instabilities in the accretion disc that fuels the BH or the clumpiness of the accreting material. Simulations following nuclear gas accretion down to sub-kpc scales \citep[e.g.][]{hopkins+quataert_2010,novak_2011,gabor+bournaud_2013} generally predict that the mass growth of the central BH predominantly happens through short, repeated accretion events, which naturally give rise to episodic bursts of AGN activity. Direct observational evidence for AGN variability comes from ionization echoes in the form of $[\OIII]$ emitting clouds (including the prototypical quasar ionization echo ``Hanny's Voorwerp'', published in \citealt{lintott_2009}; many have been found thereafter, see e.g. \citealt{keel_2012} and \citealt{schirmer_2013}), and from delayed Ly-$\alpha$ emission from nearby Lyman-$\alpha$ blobs \citep{schirmer_2016}. In both of these, recent AGN activity is required to account for the degree of ionization of the emitting gas. In the Milky Way the observed $\gamma$-ray emitting Fermi bubbles provide evidence of nuclear activity in the Galactic centre roughly $\sim 1$~Myr ago \citep[e.g.][]{su_2010,zubovas_2011}. Furthermore, AGN variability has been invoked to explain the absence of a correlation between AGN luminosity and host star formation rate as reported by a number of observational studies, despite the expected close relation between SF and AGN activity (see e.g. \citealt{alexander+hickox_2012} and \citealt{hickox_2014}, although \citealt{mcalpine_2017} argue that AGN variability is only part of the explanation). Hence, the facts that AGN are likely transient phenomena and that all galaxies are thought to harbor a BH in their centre, suggest that all galaxies are potential AGN hosts, although they are not necessarily active at the time of observation.

Rough estimates of the fraction of time that the AGN in a given galaxy is `on', also referred to as the AGN duty cycle fraction \fduty, follow from comparing the number densities of AGN and their host haloes, where the observed AGN clustering strength is used to infer the typical host halo mass \citep[see e.g.][who consider $z \simeq 2-4$]{haiman+hui_2001,martini+weinberg_2001,shen_2007}, and from comparing the time integral of the quasar luminosity function to the estimated present-day BH number density \citep[e.g.][]{yu+tremaine_2002,haiman_2004,marconi_2004}. These observational constraints typically yield $\fdutym \sim 0.1 - 10 \%$, although a related approach by \citet{shankar_2010} at $z = 3-6$ derives duty cycle fractions as high as $\fdutym \sim 10 - 90 \%$. Studies measuring the AGN occurrence in galaxies from their optical emission lines \citep[e.g.][]{kauffmann_2003,miller_2003,choi_2009} or X-ray emission \citep[e.g.][]{bongiorno_2012,mullaney_2012a} generally find that the fraction of galaxies with active AGN depends on stellar mass and redshift, as well as on the selection diagnostics used, but typical fractions range from $\sim 1 \%$ to $20 \%$ in the galaxy mass range that we consider here.

The time per `cycle' that the AGN is on, which we will refer to as the AGN lifetime, \tagn, can also be constrained observationally, using quasar proximity effects on the surrounding gas probed in absorption \citep[e.g.][]{schirber_2004,goncalves_2008,kirkman+tytler_2008,syphers+shull_2014}. Typical estimates are $\tagnm \sim 1 - 30$~Myr. However, these constraints are indirect and limited by the fact that \tagn is potentially longer than the time that the AGN has been on for now, while it is also possible that the AGN has turned off and on again since it irradiated the absorbing gas. Furthermore, based on a statistical argument, using the fraction of the X-ray detected AGN that is optically elusive and the light-travel time across the host galaxy, \citet{schawinski_2015} derived an estimate of the AGN lifetime of $\tagnm \sim 10^5$~yr.

In this work, we investigate how the fluctuating photoionizing radiation field from a central AGN affects the metal ion abundances in the CGM of the host galaxy. We mainly focus on \OVI, which is a widely studied ion in observations of quasar absorption-line systems, in particular at low redshift \citep[e.g.][]{prochaska_2011,tumlinson_2011}, but also at high redshift \citep[e.g.][]{carswell_2002,lopez_2007,turner_2015}. Observations with the Cosmic Origins Spectrograph (COS), taken as part of the COS-Halos survey, found high abundances of \OVI in the CGM of $z \sim 0.2$ star-forming galaxies, extending out to at least $150$~kpc, which is $\approx 0.5$ times the virial radius for the typical galaxy mass that was probed \citep{tumlinson_2011}. However, cosmological hydrodynamical simulations has so far not succeeded in reproducing these high \OVI columns \citep[e.g.][]{hummels_2013,ford_2016,oppenheimer_2016,suresh_2017}: they generally underpredict the observed column densities by a factor of $\approx 2 - 10$ (see e.g. \citealt{mcquinn+werk_2017} for further discussion). Here, we show that fluctuating AGN strongly enhance the \OVI in the CGM of galaxies with stellar masses of $M\sub{\ast} \sim 10^{10 - 11} \, \mathrm{M}\sub{\odot}$, both at $z = 0.1$ and at $z = 3$, and that this enhancement remains for several megayears after the central AGN fade. Hence, this provides a potential way of reconciling the predicted \OVI column densities with the observed ones. This is explored in more detail by \citet{oppenheimer_2017}.

Continuing the work by \citet{oppenheimer+schaye_2013b}, who considered a single gas pocket exposed to fluctuating AGN radiation, we here consider the CGM of galaxies selected from the Evolution and Assembly of GaLaxies and their Environments (EAGLE) simulations (\citealt{schaye_2015}, hereafter S15; \citealt{crain_2015}), where we include enhanced photoionization from a local AGN in post-processing. We follow the time-evolving abundance of circumgalactic \OVI using a reaction network \citep{oppenheimer+schaye_2013a} that captures the non-equilibrium behaviour of $133$ ions. To quantify to what extent AGN proximity zone fossils affect the interpretation of CGM column density measurements from quasar absorption-line systems, we present predictions of the average enhancement of the \OVI column density and covering fraction in a fluctuating AGN radiation field. Furthermore, we calculate the probability of observing a significant AGN fossil effect\footnote{We note that the `fossil effect' that we refer to in this work, includes the effects from both the finite light-travel time (i.e. ionization echoes) and from delayed recombination after the enhanced incident radiation ceases.} (i.e. a CGM \OVI column density that is out of equilibrium by at least $0.1$~dex), while the central AGN in the galaxy is inactive. This gives an indication of the fraction of quasar absorption-line systems that are likely affected by AGN fossil effects. We explore the dependence on impact parameter, galaxy stellar mass and redshift, as well as the dependence on the adopted parameters used to model the fluctuating AGN: we vary the AGN luminosity (by varying the Eddington ratio, \eddratio, where $L\sub{Edd}$ is the Eddington luminosity), lifetime and duty cycle fraction.

This paper is organized as follows. In Section~\ref{sec:methods}, we describe the simulation output used, the AGN model that we implement in post-processing and our method for calculating the time-variable column densities of CGM ions. We also introduce the three quantities we use to quantify the significance of the AGN fossil effect. In Section~\ref{sec:results}, we present our results for \OVI and show how they depend on the properties of the galaxy and the adopted AGN model parameters. We briefly present results for other metal ions in Section~\ref{sec:other_metals} and we summarize our findings in Section~\ref{sec:conclusions}.


\section{Methods}
\label{sec:methods}

We begin by giving a brief overview of the EAGLE simulation code and the non-equilibrium ionization module, followed by a description of the fluctuating AGN model used to photoionize the CGM of the selected galaxies. We then describe how we calculate column densities from the ion abundances predicted by the simulation, and how we quantify the significance of the AGN fossil effects.

\subsection{EAGLE simulations}
\label{sec:simulations}

The EAGLE simulations were run with a heavily modified version of the smoothed particle hydrodynamics (SPH) code \textsc{Gadget3} \citep[last described by][]{springel_2005}. A collection of updates, referred to as \textsc{Anarchy} (Appendix A of S15; see also \citealt{schaller_2015}), has been implemented into the code, including the use of a pressure-entropy formulation of SPH \citep{hopkins_2013}. The adopted cosmological parameters are taken from \citet{planck_2014}: $\left[ \Omega\sub{m},\Omega\sub{b},\Omega\sub{\Lambda},\sigma\sub{8},n\sub{s},h \right]=\left[ 0.307,0.04825,0.693,0.8288,0.9611,0.6777 \right]$.

The implemented subgrid physics is described in detail in S15. In brief, star formation is modelled as the stochastic conversion of gas particles into star particles, following the pressure-dependent prescription of \citet{schaye+dallavecchia_2008} in combination with a metallicity-dependent density threshold \citep[given by][]{schaye_2004}. Because the simulations do not model a cold phase, a global temperature floor, corresponding to the equation of state $P \propto \rho\up{4/3}$ and normalized to $8000$~K at a density of $n\sub{H} = 0.1$~cm$\up{-3}$, is imposed on the gas in the interstellar medium. When computing the ionization balance (Section~\ref{sec:network}), we set the temperature of star-forming gas to $T = 10^4$~K, as its temperature given in the simulation merely reflects an effective pressure due to the imposed temperature floor.

Star particles enrich their surroundings through the release of mass and metals in stellar winds and supernova explosions (Type Ia and Type II) according to the prescriptions of \citet{wiersma_2009b}. The adopted stellar initial mass function is taken from \citet{chabrier_2003}. During the course of the simulation, the abundances of $11$ elements (i.e. H, He, C, N, O, Ne, Mg, Si, Fe, Ca and Si) are followed, which are used to calculate the equilibrium rates of radiative cooling and heating in the presence of cosmic microwave background and \citet[][HM01]{haardt+madau_2001} UV and X-ray background radiation \citep{wiersma_2009a}. The time-dependent abundances of $133$ ion species are calculated in post-processing, as we describe in Section~\ref{sec:network}, and are not used for the cooling and heating rates.

Energy feedback from star formation and AGN is implemented by stochastically heating gas particles surrounding newly formed star particles and BH particles, respectively \citep{dallavecchia+schaye_2012}. The BHs, with which haloes are seeded as in \citet{springel_2005}, grow through mergers and gas accretion, where the accretion rate takes into account the angular momentum of the gas \citep[][S15]{rosasguevara_2015}. The subgrid parameters in the models for stellar and AGN feedback have been calibrated to reproduce the observed present-day galaxy stellar mass function, the sizes of galaxies, and the relation between stellar mass and BH mass.

In this work, we focus on four galaxies selected from the EAGLE reference simulation. This simulation (referred to as \emph{Ref-L100N1504} in S15) was run in a periodic, cubic volume of $L = 100$ comoving Mpc on a side. It contains $N = 1504^3$ dark matter particles and an equal number of baryonic particles with (initial) masses of $m\sub{dm}=9.7 \times 10^6\ {\rm M}\sub{\odot}$ and $m\sub{b}=1.8 \times 10^6\ {\rm M}\sub{\odot}$, respectively, and with a gravitational softening length of $2.66$ comoving kpc, limited to a maximum physical scale of $0.7$ proper kpc.

Haloes and galaxies are identified from the simulation using the Friends-of-Friends and \textsc{Subfind} algorithms \citep{dolag_2009}. Galaxies are subdivided into `centrals' and `satellites', where the former are the galaxies residing at the minimum of the halo potential. The mass of the halo, referred to as the virial mass $M\sub{vir}$, is defined as the mass enclosed within a spherical region centred on the minimum potential, within which the mean density equals $200$ times the critical density of the Universe. The corresponding virial radius and temperature are denoted by $R\sub{vir}$ and $T\sub{vir}$, respectively.


\subsection{Non-equilibrium ionization module}
\label{sec:network}

To model the time-variable abundances of ion species in the CGM of our simulated galaxies, we use the reaction network introduced by \citet{oppenheimer+schaye_2013a}. It follows the $133$ ionization states of the $11$ elements that are used to compute the (equilibrium) cooling rates in the simulation, as well as the number density of electrons. The reactions included in the network are those corresponding to radiative and di-electric recombination, collisional ionization, photoionization, Auger ionization and charge transfer. Given the set of reaction rates, the module calculates the ionization balance as a function of time, without making the assumption that the gas is in ionization equilibrium. While it is possible to integrate the module into the simulation and to calculate ion abundances and ion-by-ion cooling rates on the fly \citep{richings+schaye_2016,oppenheimer_2016}, we here work strictly in post-processing. This means that we do not include dynamical evolution or evolution of the temperature when we solve for the ionization state of the gas. We note that, in contrast to the cooling rates, which are calculated from `kernel-smoothed' element abundances \citep[i.e. the ratio of the mass density of an element to the total mass density per particle;][]{wiersma_2009b}, we use particle-based element and ion abundances (i.e. the fraction of the mass in an element or ion) in the reaction network.

The non-equilibrium ionization module enables us to explore the effect on the CGM of a time-variable source of ionizing radiation, in our case of an AGN positioned in the centre of the galaxy. A source with specific intensity $f_{\nu}$ photoionizes ions of atomic species $x$ from state $i$ to $i+1$ at a rate
\begin{equation}\label{eq:phot_rate}
\Gamma_{x_i\mathrm{,AGN}} = \int_{\nu_{0,x_i}}^{\infty} \frac{f_v}{h\nu} \sigma_{x_i}(\nu) \, \D \nu,
\end{equation}
where $\nu$ is the frequency, $\nu_{0,x_i}$ is the ionization frequency, $h$ is the Planck constant and $\sigma_{x_i}(\nu)$ is the photoionization cross-section. The evolution of the number density $n_{x_i}$ of ions in state $x_i$ is then given by
\begin{multline}\label{eq:n_evo}
\frac{\D n_{x_i}}{\D t} = n_{x_{i+1}} \alpha_{x_{i+1}} n\sub{e} + n_{x_{i-1}} \left( \beta_{x_{i-1}} n\sub{e} + \Gamma_{x_{i-1}\mathrm{,EGB}} \right. \\
\left. + \Gamma_{x_{i-1}\mathrm{,AGN}} \right) - n_{x_i} \left( \left( \alpha_{x_i} + \beta_{x_i} \right) n\sub{e} + \Gamma_{x_i\mathrm{,EGB}} + \Gamma_{x_i\mathrm{,AGN}} \right),
\end{multline}
where charge transfer and Auger ionization have been omitted from the equation for simplicity. Here $n\sub{e}$ is the free electron number density, which depends mostly on the abundance and ionization state of hydrogen. $\alpha_{x_i}$ and $\beta_{x_i}$ are the rates of recombination (including both radiative and di-electric) and collisional ionization, respectively, which depend on the local temperature. The photoionization rate from the extra-galactic background, $\Gamma_{x_i\mathrm{,EGB}}$, is calculated from equation~\ref{eq:phot_rate} using the redshift-dependent HM01 spectral shape, consistent with the background radiation included in the simulation.


\subsection{Ion column densities}
\label{sec:calc_coldens}

We compute column densities ($N$) of ions in the CGM by projecting a cylindrical region with a radius of $2R\sub{vir}$ and a line-of-sight length of $2$~Mpc, centred on the centre of the galaxy, onto a 2D grid of $1000 \times 1000$ pixels\footnote{We have checked that the number of grid pixels is sufficiently high so that the CGM column densities are converged.}. For each grid pixel, we calculate the ion column densities from the particle ion abundances using two-dimensional, mass-conserving SPH interpolation. Throughout this work, we will mainly focus on \OVI. We therefore define the quantities we use to quantify the significance of the AGN fossil effect specifically for \OVI. However, these quantities are defined for other ions in a similar way.

We consider the column density of circumgalactic \OVI up to impact parameters (i.e. projected galactocentric distances) of $2R\sub{vir}$. To construct column density profiles, which we denote by $\Nm{OVI}(R)$, we take the median column density of all the grid pixels within an impact parameter range centred on $R$. We take the median, rather than the mean or the total number of ions divided by the area of the bin, since this mimics the cross-section-weighted observations of column densities in quasar absorption-line studies more closely.


\subsection{Galaxy sample}
\label{sec:galaxies}

\begin{figure*}
\begin{center}
\includegraphics[width=\textwidth]{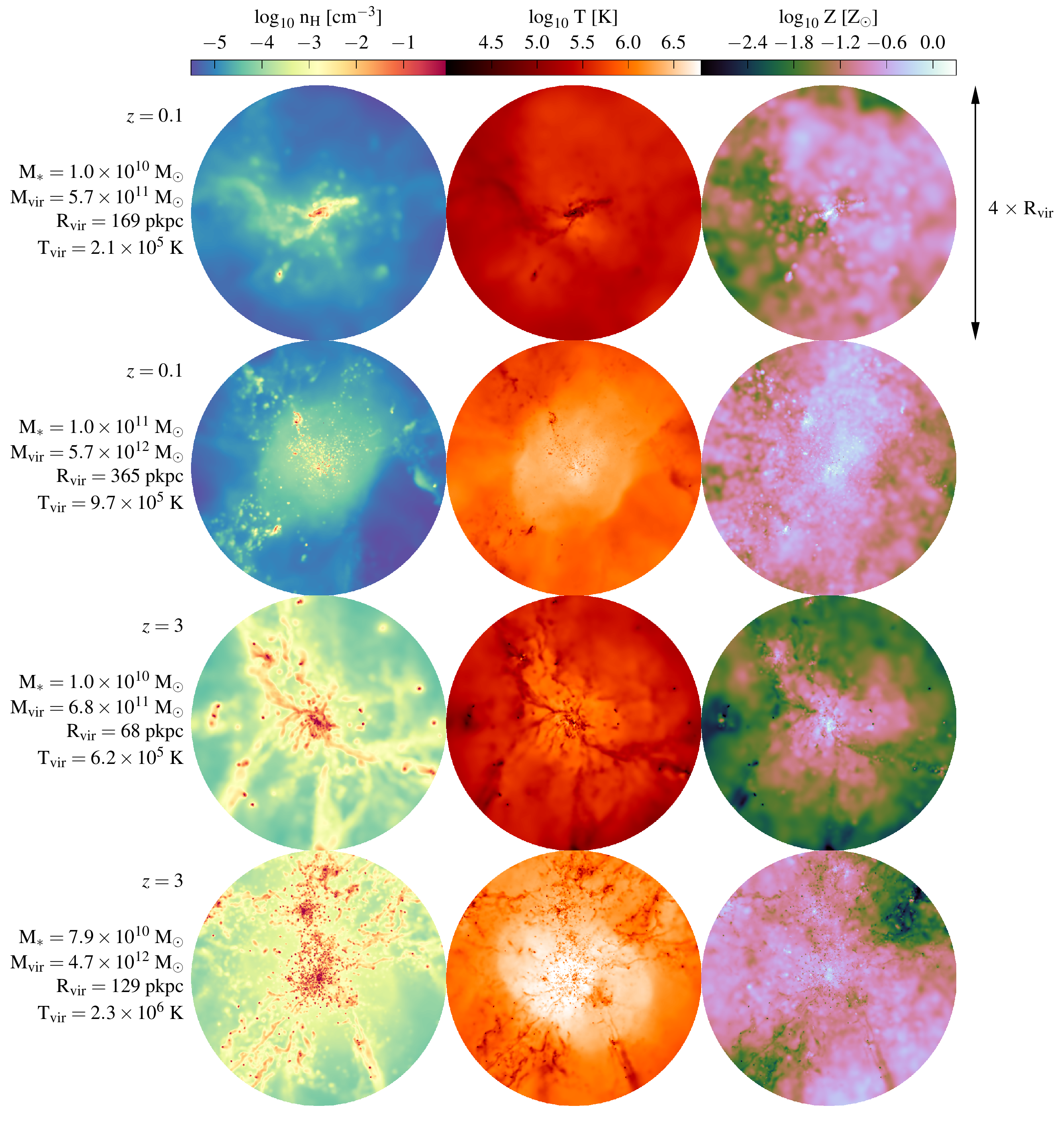}
\end{center}
\caption{Maps of the hydrogen number density (left), temperature (middle) and metallicity (right; normalized to the solar metal mass fraction $Z\sub{\odot} = 0.0129$) for the CGM of the four galaxies considered in this work. These are all central galaxies and have been selected from the EAGLE \emph{Ref-L100N1504} simulation. Their stellar mass (\mstarsim{10} and \mstarsim{11}), redshift ($z = 0.1$ and $z = 3$) and virial properties are listed on the left. The colour-coding indicates the mass-weighted quantity projected onto a $2$D grid with radius $2R\sub{vir}$ using SPH interpolation, within a slice of $2$~Mpc thickness centred on the galaxy.}
\label{fig:galaxies}
\end{figure*}

To explore how the strength of AGN proximity zone fossils depends on galaxy mass and redshift, we consider two (central) galaxies with stellar masses of \mstarsim{10} and \mstarsim{11} at $z = 3$ and two galaxies with similar stellar masses at $z = 0.1$\footnote{These correspond to the galaxies with $\textsc{GalaxyID} = 19523883, 18645002, 10184330, 15484683$ in the publicly available EAGLE catalogue at \url{http://www.eaglesim.org/database.php} \citep{mcalpine_2016}.}. These galaxies have been selected to be `representative' galaxies, with stellar-to-halo mass ratios that are close to the mean and median value at the respective stellar mass and redshift. Fig.~\ref{fig:galaxies} shows maps of their hydrogen number density (left column), temperature (middle column) and metallicity (right column). These maps have been made by projecting a cylindrical region with a radius of $2R\sub{vir}$ and a length of $2$~Mpc, centred on the galaxy, onto a $2$D grid (similarly to how we compute ion column densities; see Section~\ref{sec:calc_coldens}) and calculating the mass-weighted quantity in each grid cell using SPH interpolation. The stellar masses, halo masses, virial radii and virial temperatures of the galaxies are listed on the left. The most evident difference between $z=0.1$ and $z=3$ is the higher density of the CGM at high redshift, with the galaxies being more embedded in filamentary structures.

\begin{figure*}
\begin{center}
\includegraphics[width=\textwidth]{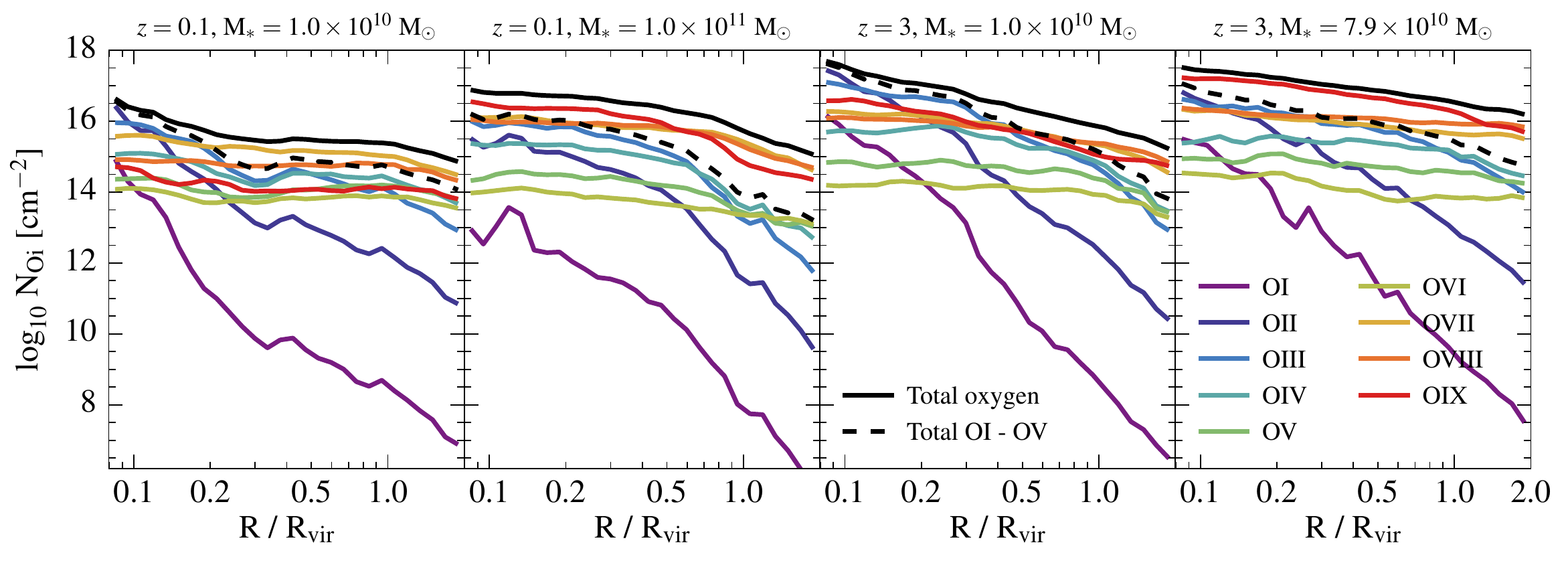}
\end{center}
\caption{Equilibrium column density profiles of oxygen ions for, from left to right, the \mstarsim{10} and \mstarsim{11} galaxies at $z = 0.1$, and the \mstarsim{10} and \mstarsim{11} galaxies at $z = 3$. The coloured curves show the individual ion column densities as a function of impact parameter, given by the median column density in logarithmic impact parameter intervals between $R / R\sub{vir} = 0.08$ and $R / R\sub{vir} = 2.0$. The black, solid (dashed) curves show the total column density of oxygen (of ion states \OI to \OV). Most of the oxygen resides in the high ion states (mostly in \OVII{ }- \OVIII for the galaxies at $z = 0.1$, and in \OVIII{ }- \OIX at $z = 3$). The \OVI state is always subdominant.}
\label{fig:eq_ab_O}
\end{figure*}

Without any AGN proximity effects, the column density profiles of the different oxygen ions in the CGM of the four galaxies are as given in Fig.~\ref{fig:eq_ab_O}. In general, the ionization state of the gas increases with increasing impact parameter: the column densities of the lower-state ions (\OI{ }- \OV) decrease significantly, while the profiles of the higher-state ions are flatter. This is related to the fact that the density (and hence, the recombination rate) is lower at larger galactocentric radii, while the gas still receives the same background radiation. At a fixed $R / R\sub{vir}$, the ionization state is higher for more massive galaxies, owing to their higher CGM temperatures (see Fig.~\ref{fig:galaxies}).

Evident for all four galaxies is that the column density of \OVI is relatively low compared to the column densities of the other oxygen ions. The dominant oxygen state is generally \OVII{ }- \OVIII for the galaxies at low redshift, and \OVIII{ }- \OIX at high redshift. As was e.g. pointed out by \citet{oppenheimer_2016}, \OVI is only the tip of the iceberg of the CGM oxygen content. Since the ion fraction of \OVI in collisional ionization equilibrium peaks at $T\sub{peak} \sim 10^{5.5}$~K, where gas cooling is fast, significant quantities of collisionally ionized \OVI only exist if the virial temperature of the halo is close to $T\sub{peak}$. Otherwise, gas predominantly exists at $T < 10^5$~K or at $T > 10^6$~K, where the ion fraction is lower, which is why \N{OVI} in the CGM of $M\sub{vir} \gtrsim 10^{12} \, \mathrm{M}\sub{\odot}$ galaxies decreases with increasing halo mass \citep[see fig. 4 of][]{oppenheimer_2016}. The photoionized phase of \OVI arises at $T < 10^5$~K and at lower densities than the collisionally ionized phase. Therefore, the CGM of low-mass galaxies, with $T\sub{vir} \ll T\sub{peak}$, exhibits a significant fraction of gas in a temperature and density regime where the ion fraction of \OVI is also high. However, if the galaxy stellar mass is low, the metallicity and total mass in oxygen are also low, which results in a low \N{OVI} despite the high ion fraction.

The galaxies with \mstarsim{10} considered in this work have virial temperatures that are close to $T\sub{peak}$ (somewhat lower for the one at $z = 0.1$ and somewhat higher for the one at $z = 3$), while the galaxies with \mstarsim{11} have $3 - 7$ times higher virial temperatures than $T\sub{peak}$. This means that especially at small radial distances from the high-mass galaxies, the \OVI is mostly collisionally ionized. At larger distances, in particular for the low-mass galaxies, an increasing fraction of the \OVI is photoionized (see Section~\ref{sec:results}).


\subsection{AGN model}
\label{sec:agn_model}

\begin{figure}
\begin{center}
\includegraphics[width=\columnwidth]{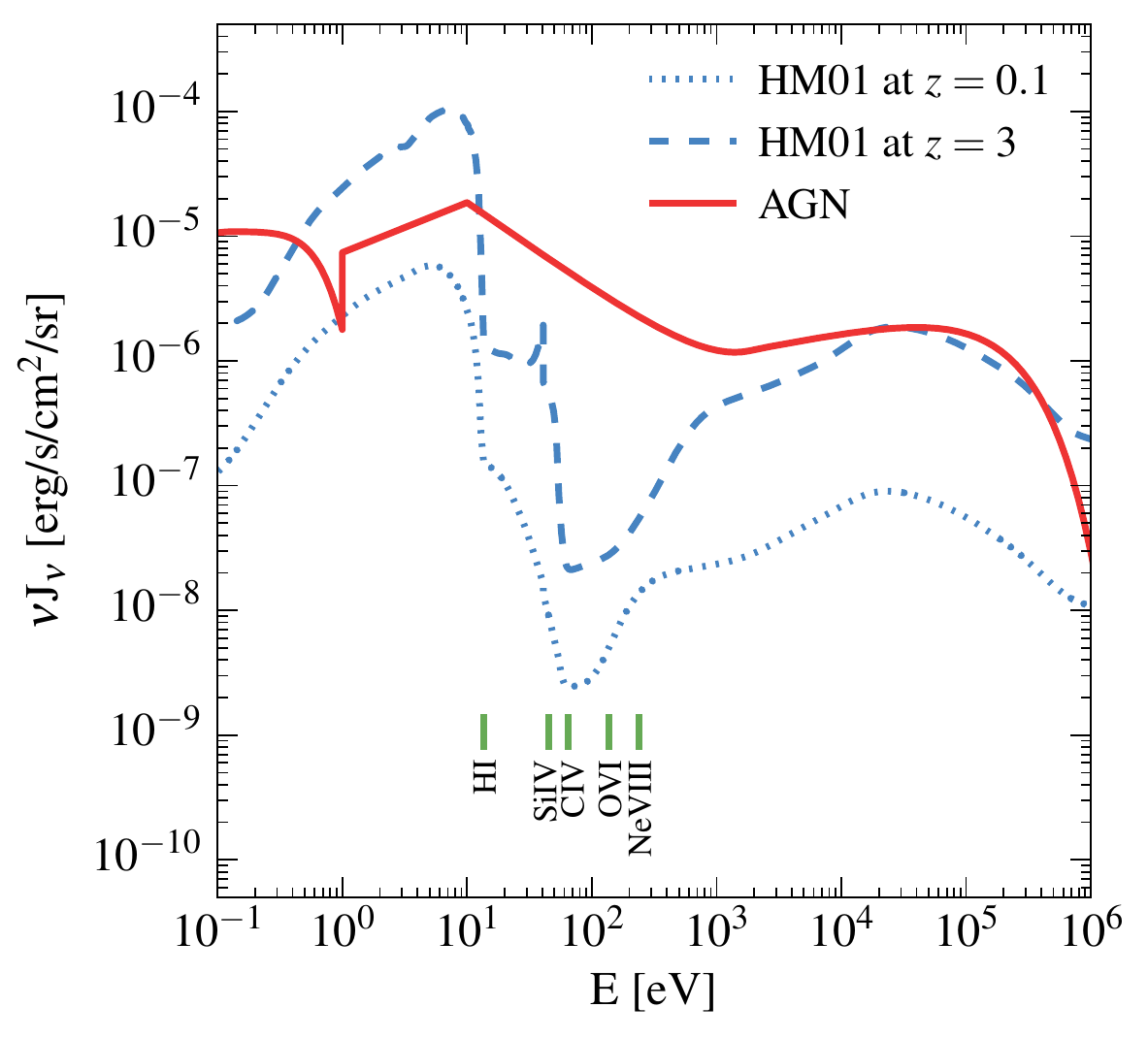}
\end{center}
\caption{The model spectrum for the homogenous UV background at $z = 0.1$ (blue, dotted line) and $z = 3$ (blue, dashed line), adopted from \citet{haardt+madau_2001}, and the model spectrum for the AGN (red, solid line), adopted from \citeauthor{sazonov_2004} (\citeyear{sazonov_2004}; i.e. the `unobscured' quasar model spectrum). Note that $f\sub{\nu} = 4\pi J\sub{\nu}$. The normalization of the AGN spectrum corresponds to an AGN with $\eddratiom = 1$, where $M\sub{BH} = 10^{7} \, \mathrm{M}\sub{\odot}$, at a distance of $100$ pkpc. This is equivalent to a bolometric luminosity of $L = 1.3 \times 10^{45}$~erg/s and a strength of $J\sub{\nu} = 10^{-20.3}$~erg/s/cm$^2$/Hz/sr at $E = 1$~Ryd. The ionization energies of a few commonly observed metal ions are indicated at the bottom.}
\label{fig:agnspec}
\end{figure}

\begin{table}
\centering
\caption{Parameter values for the AGN model explored in this work: the AGN luminosity as a fraction of the Eddington luminosity (\eddratio), the AGN lifetime per cycle (\tagn), and the fraction of time that the AGN is on (\fduty).}
\begin{tabular}{l r}
\hline
Parameter & Values\\
\hline
\eddratio & $0.01, 0.1, 1.0$\\
\tagn & $10^5, 10^6, 10^7$~yr\\
\fduty & $1, 2, 5, 10, 20, 50 \%$\\
\hline
\end{tabular}
\label{tab:par_values}
\end{table}

Having selected our four galaxies, we include a variable photoionizing radiation field in post-processing as follows. We assume that the radiation source is located at the minimum of the potential of the galaxy (including its subhalo), that irradiates the gas isotropically with a certain luminosity, spectral shape and periodicity. The ionizing radiation propagates through the galaxy and CGM with the speed of light\footnote{Note that while we account for the finite light-travel time of the AGN radiation through the CGM, we do not consider the differential light-travel times from different parts of the CGM to the observer.}, where the spatial position, density, and temperature of the gas have been fixed to those output by the simulation at the respective redshift. Note that this means that we do not include the effect of photoheating by the local AGN. However, \citet{oppenheimer+schaye_2013b} show that the change in the temperature due to photoheating is generally small (e.g. $\Delta \log\sub{10} T \lesssim 0.1$~dex at $100$~kpc from an AGN that is comparable to a local Seyfert). In Appendix~\ref{sec:photoheating}, we explicitly show for our set-up that the effect of photoheating on the \OVI abundance of the CGM is expected to be small compared to the effect of photoionization.

We assume that the gas in and around the galaxy is optically thin, as we only consider high-ionization state ions, which occur in low-density CGM gas where self-shielding against ionizing radiation is unimportant. It is, however, possible that optically thick structures are present near the centre of the galaxy, in the form of a dusty torus surrounding the BH or dense gas in clumps or in the galactic disc, that would make the radiation field from the AGN anisotropic. \citet{oppenheimer_2017} explore the AGN fossil effect in an anisotropic radiation field, using a bicone model with $120\degr$ opening angles, mimicking an obscuring nuclear torus either aligned with the galactic rotation axis or at a random orientation: they find that, even though only half of the CGM volume is irradiated at each AGN episode, the fossil effect is more than half as strong as in the isotropic case. This is because, on the one hand, a $2\pi$ steradian solid angle still affects the majority of the sightlines through the CGM, and, on the other hand, because the AGN eventually ionizes more than half of the CGM volume as the cone direction varies with time, as a result of the significant recombination timescales of the metal ions. Any other obscuring structures, in the galactic disc or in isolated clumps, likely cover a much smaller solid angle, so we expect their effect on the strength of the fossil effect to be small. Moreover, anisotropic AGN radiation would require larger duty cycle fractions for the same observed quasar luminosity function, which reduces (and perhaps compensates entirely for) any effects of anisotropic radiation, as larger duty cycle fractions tend to increase the strength of the fossil effect (see Section~\ref{sec:cycle_dep}).

Switching the AGN on or off happens instantaneously (i.e. the AGN is either off or at a fixed luminosity). As soon as the ionization front reaches a gas parcel, the AGN flux is added to the uniform HM01 background flux. Fig.~\ref{fig:agnspec} shows a comparison between the HM01 spectrum (at $z=0.1$ and $z=3$) and the AGN spectral shape, which we adopt from \citet{sazonov_2004}. In this work, we explore variations of the AGN Eddington ratio \eddratio, lifetime \tagn and duty cycle fraction \fduty, where we base our choices of these parameters on observational constraints compiled from the literature. The parameter values we explore are listed in Table~\ref{tab:par_values}.

We consider Eddington ratios of $0.01$, $0.1$ and $1.0$, which we convert into a (bolometric) luminosity using the standard expression for the Eddington luminosity,
\begin{equation}\label{eq:edd_lum}
L\sub{Edd} = \frac{4 \pi G m\sub{p} c}{\sigma\sub{T}} M\sub{BH}.
\end{equation}
Here the $G$ is the gravitational constant, $m\sub{p}$ is the proton mass, $c$ is the speed of light and $\sigma\sub{T}$ is the Thomson scattering cross-section. We fix the mass of the BH, $M\sub{BH}$, to $M\sub{BH} = 10^{-3} M\sub{\ast}$ at both $z=0.1$ and $z=3$\footnote{We adopt $M\sub{BH} = 10^{-3} M\sub{\ast}$ to calculate $L\sub{Edd}$, rather than the BH mass from the simulation, in order to have an AGN luminosity that is representative for the whole galaxy population at the given redshift and stellar mass. In this way, $L\sub{Edd}$ is insensitive to the deviation of the simulated $M\sub{BH}$ from the median $M\sub{BH}(M\sub{\ast})$ relation for the four galaxies considered in this work.}, which is approximately the local relation observed over a wide range of galaxy masses \citep[e.g.][]{merritt+ferrarese_2001,marconi+hunt_2003,haring+rix_2004}. There are, however, indications that the normalization increases with increasing redshift (see e.g. \citealt{salviander+shields_2013} or fig. 38 of \citealt{kormendy+ho_2013}; but see \citealt{sun_2015}), but it remains uncertain to what extent. Note that our exploration of different Eddington ratios can be interpreted as varying the $M\sub{BH}(M\sub{\ast})$ relation (or both the \eddratio and the $M\sub{BH}(M\sub{\ast})$ relation). At high redshift ($0.5 \lesssim z < 4-5$), AGN are often found to exhibit near-Eddington luminosities, with narrow (width $\lesssim 0.3$~dex) \eddratio distributions, typically peaking in between $0.1$ and $1.0$ \citep[e.g.][]{kollmeier_2006,netzer_2007,shen_2008}. At low redshift ($z \lesssim 0.3$), however, observations of \eddratio find distributions that are wider and that span values significantly lower than $1$ \citep[typically $\lesssim 0.1$; see e.g.][]{greene+ho_2007,heckman_2004,kauffmann+heckman_2009}. Hence, we adopt default values of $\eddratiom = 0.1$ at $z = 0.1$ and $\eddratiom = 1$ at $z = 3$, when we compare galaxies at different redshifts. We investigate the impact of adopting a higher or a lower \eddratio for the \mstarsim{11} galaxy at $z=0.1$ and the \mstarsim{10} galaxy at $z=3$ in Section~\ref{sec:lum_dep}.

Since the EAGLE simulations lack the resolution to make reliable predictions on the periodicity of nuclear gas accretion, we rely on observations for constraints on the AGN lifetime and duty cycle fraction. Statistical arguments and observations of individual absorption systems and Ly$-\alpha$ emitters near bright quasars constrain the typical AGN lifetime to $\tagnm = 10^5 - 10^7$~yr (see Section~\ref{sec:introduction} for references). Estimates of the AGN duty cycle fraction, which are generally derived from the fraction of a sample of galaxies hosting active AGN, also span a large range of values: they range from less than $1 \%$ to as high as $90 \%$ (see Section~\ref{sec:introduction}). Hence, we explore duty cycles of $\fdutym = 1, 2, 5, 10, 20, 50 \%$. We refer to \tagn as the `AGN-on' time and to the time in between two subsequent AGN-on phases as the `AGN-off' time ($t\sub{off}$). We refer to the sum of one AGN-on phase and one AGN-off phase as one full AGN cycle:
\begin{equation}\label{eq:t_cycle}
t\sub{cycle} = \tagnm + t\sub{off} = \tagnm \frac{100 \%}{\fdutym}.
\end{equation}


\subsection{Quantifying the AGN fossil effect}
\label{sec:quant_agn_fossil}

The imprint on the column densities of CGM ions of past AGN activity after the AGN has faded, is what characterizes an AGN proximity zone fossil. We quantify the fossil effect for \OVI by measuring the (logarithmic) difference between the current \OVI column density and its initial value in ionization equilibrium, $\Nm{OVI}\up{t=0}$. For example, to explore the spatial variation of the fossil effect at a given timestep, we calculate
\begin{equation}\label{eq:delta_N}
\deltaNm{OVI} \equiv \log\sub{10} \left( \Nm{OVI} \middle/ \mathrm{cm}^{-2} \right) - \log\sub{10} \left( \Nm{OVI}\up{t=0} \middle/ \mathrm{cm}^{-2} \right)
\end{equation}
at every pixel of the projection grid.

To quantify the significance of the AGN fossil effect in a statistical way and to enable a comparison between different AGN set-ups, we consider:
\begin{itemize}
\item $7$ logarithmic impact parameter bins of width $0.2$~dex between $0.08R\sub{vir}$ and $2R\sub{vir}$, in which we take the median column density of all grid pixels (as described in Section~\ref{sec:calc_coldens}) to obtain $\log\sub{10} \Nm{OVI}(R)$;
\item the time average of $\log\sub{10} \Nm{OVI}(R)$ during the AGN-off time, i.e. in between two AGN-on phases. For combinations of AGN model parameters for which the fossil effect accumulates over multiple cycles (i.e. short \tagn and large \fduty; see Section~\ref{sec:cycle_dep}), we calculate the average of $\log\sub{10} \Nm{OVI}(R)$ over $t\sub{off}$ after the fluctuating $\log\sub{10} \Nm{OVI}(R)$ has reached an asymptotic value, reflecting a net balance between the number of ionizations and recombinations per cycle.
\end{itemize}
For each galaxy and AGN set-up, this yields a single quantity as a function of impact parameter,
\begin{equation}\label{eq:av_N}
\avNm{OVI} \equiv \frac{1}{t\sub{cycle} - t\sub{AGN}} \int_{t\sub{AGN}}^{t\sub{cycle}} \log\sub{10} \Nm{OVI}(R) \, \D t,
\end{equation}
that can be compared to the corresponding value in equilibrium.

Another commonly measured quantity in studies of CGM ion abundances, is the ion covering fraction. We define the \OVI covering fraction, $\fcovm{OVI}(R)$, as the fraction of the pixels within the impact parameter range around $R$ that have $\Nm{OVI} > 10^{14.0} \, \mathrm{cm}^{-2}$. Similarly to the average column density, we calculate its average over the AGN-off time:
\begin{equation}\label{eq:av_f_cov}
\avfcovm{OVI} \equiv \frac{1}{t\sub{cycle} - t\sub{AGN}} \int_{t\sub{AGN}}^{t\sub{cycle}} \fcovm{OVI}(R) \, \D t.
\end{equation}

While \avN{OVI} and \avfcov{OVI} characterize the \emph{strength} of the fossil effect averaged over time, we define one additional quantity to indicate the \emph{probability} of observing a significant AGN fossil effect while the AGN is off. We calculate the fraction of the time in between two AGN-on phases for which $\log\sub{10} \Nm{OVI}(R)$ is offset from equilibrium by at least $0.1$~dex. This again is a function of impact parameter, and allows a comparison between different galaxies and AGN set-ups.


\section{Results for OVI}
\label{sec:results}

\begin{figure*}
\begin{center}
\includegraphics[width=\textwidth]{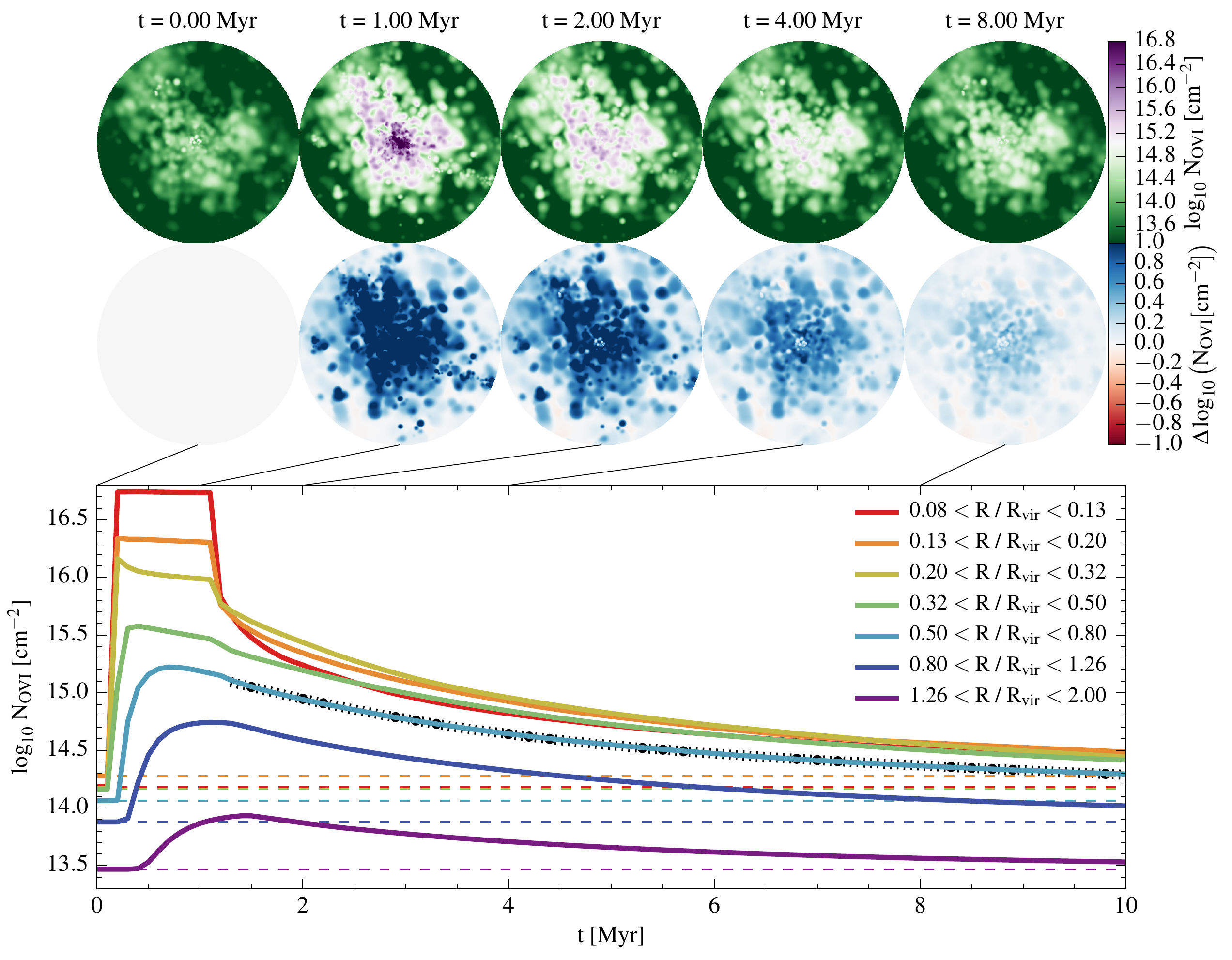}
\end{center}
\caption{The evolution of the \OVI column density around the \mstar{1.0}{10} galaxy at $z = 3$, for an $\eddratiom = 1.0$ AGN that is on for $1$~Myr and off for $9$~Myr (i.e. $\tagnm = 10^6$~yr and $\fdutym = 10 \%$). The maps show the \OVI column density (upper row) and difference in $\log\sub{10} \Nm{OVI}$ with respect to $t = 0$~Myr (equation~\ref{eq:delta_N}; lower row) at $t = 0, 1, 2, 4, 8$~Myr. The bottom panel shows the evolution of the median \N{OVI} (solid lines), as well as the equilibrium value at $t = 0$~Myr (dashed lines), in $7$ logarithmic impact parameter intervals between $0.08R\sub{vir}$ and $2R\sub{vir}$. The fit to $\Nm{OVI}(R)$ at $0.5 < R / R\sub{vir} < 0.8$ (black, dotted line) shows that the evolution of $\Nm{OVI}(R)$ after the AGN turns off is well-approximated by a sum of two exponential functions (equation~\ref{eq:two_exp}).}
\label{fig:example_seq_OVI}
\end{figure*}

Prior to exploring the dependence of AGN fossil effects on the impact parameter (Section~\ref{sec:rad_dep}), the stellar mass and redshift of the galaxy (Section~\ref{sec:mass_dep}) and the strength, lifetime, and duty cycle of the AGN (Section~\ref{sec:agn_dep}), we will show how the column density of circumgalactic \OVI changes as a function of time for one particular set of AGN model parameters. We focus here on the \mstar{1.0}{10} galaxy at $z = 3$.

The maps at the top of Fig.~\ref{fig:example_seq_OVI} show the \OVI column density (upper row) at $t = 0, 1, 2, 4, 8$~Myr, for an AGN with an Eddington ratio of $\eddratiom = 1.0$ that is on for $1$~Myr and off for $9$~Myr (i.e. $\tagnm = 10^6$~yr and $\fdutym = 10 \%$). The maps in the lower row show the difference in $\log\sub{10} \Nm{OVI}$, \deltaN{OVI} (equation~\ref{eq:delta_N}), with respect to the equilibrium value at $t = 0$~Myr. As in Fig.~\ref{fig:galaxies}, all maps show the circumgalactic gas up to impact parameters of $2 R\sub{vir}$. From $t = 0$~Myr to $t = 1$~Myr, the enhanced radiation field from the AGN ionizes a significant fraction of the lower-state oxygen ions to \OVI, leading to a large increase in the column density. After $t = 1$~Myr, when the AGN switches off and the radiation field returns instantaneously to the uniform HM01 background, this \OVI enhancement starts decreasing again. However, due to the significant recombination times of oxygen ions, and the series of ions that the oxygen needs to recombine through, the gas is left in an overionized state for several megayears. This remnant of past AGN activity in which ionization equilibrium has not been achieved yet, is what characterizes an AGN proximity zone fossil. The fossil effect is illustrated more quantitatively in the bottom panel of the figure, which shows the evolution of the median \OVI column density in $7$ impact parameter bins (solid lines). Naturally, the AGN-induced boost in \N{OVI} with respect to the equilibrium value (dashed lines) is stronger at smaller galactocentric distances\footnote{Note that the short time delay in the increase and decrease of \N{OVI} is due to the light-travel time of the ionization front.}: at $R \sim R\sub{vir}$ the boost is about $0.8$~dex, while for $R \lesssim 0.5 R\sub{vir}$ it is $\gtrsim 1.4$~dex. Except in the outer two bins, \N{OVI} even slightly decreases again during the AGN-on time, as \OVI is ionized to higher states.

After the AGN turns off, the timescale on which \N{OVI} returns to equilibrium depends mostly on the recombination time of \OVI to \OV, $t\sub{rec}\up{OVI}$, and the recombination time of \OVII to \OVI, $t\sub{rec}\up{OVII}$. The latter is important as it is associated with the recombination of higher-state oxygen ions to \OVI, $t\sub{rec}\up{OVII}$ being the bottleneck in this recombination sequence. For $t > \tagnm$ ($+$ the radius-dependent time delay), when the gas is left in an overionized state, the evolution of the surplus of \OVI number density can be approximated as a combination of two recombination processes:
\begin{equation}\label{eq:n_evo_approx}
\frac{\D n\sub{OVI}}{\D t} = n\sub{OVII} \alpha\sub{OVII} n\sub{e} - n\sub{OVI} \alpha\sub{OVI} n\sub{e}.
\end{equation}
For fixed values of $\alpha\sub{OVI}$, $\alpha\sub{OVII}$ and $n\sub{e}$ the solution to this differential equation is a sum of two exponential functions,
\begin{equation}\label{eq:two_exp}
n\sub{OVI}(t) = C\sub{1}\mathrm{e}^{-\alpha\sub{OVI} n\sub{e} t} + C\sub{2}\mathrm{e}^{-\alpha\sub{OVII} n\sub{e} t},
\end{equation}
where $C\sub{1}$ and $C\sub{2}$ are normalization constants. The exponential decay rates are related to the recombination timescales as $t\sub{rec}\up{OVI} = 1 / (\alpha\sub{OVI} n\sub{e})$ and $t\sub{rec}\up{OVII} = 1 / (\alpha\sub{OVII} n\sub{e})$, which describe the evolution of $n\sub{OVI}$ on short and long timescales, respectively. We find that, even though equation~(\ref{eq:two_exp}) describes the evolution of the \OVI number density, the evolution of $\Nm{OVI}(R)$ after AGN turn-off can also be approximated by a sum of two exponentials. We show the fit (performed in logarithmic space) for $\Nm{OVI}(R)$ and $0.5 < R / R\sub{vir} < 0.8$ (black, dotted line) in Fig.~\ref{fig:example_seq_OVI} to illustrate this. The best-fitting $t\sub{rec}\up{OVI}$ and $t\sub{rec}\up{OVII}$ then give us an indication of the effective re-equilibration timescales of \OVI: we find $t\sub{rec}\up{OVI} = 1.4$~Myr and $t\sub{rec}\up{OVII} = 12.1$~Myr at $0.5 < R / R\sub{vir} < 0.8$, which are similar to the expected recombination timescales in $n\sub{H} \sim 10^{-3.5}$~cm$^{-3}$ and $T \sim 10^{4.5}$~K gas. However, in reality $t\sub{rec}\up{OVI}$ and $t\sub{rec}\up{OVII}$ are not constants: they depend on the local temperature and density (and on the ionization state of hydrogen through $n\sub{e}$). Since the gas in a certain impact parameter range spans a range of densities and temperatures, the best-fit $t\sub{rec}\up{OVI}$ and $t\sub{rec}\up{OVII}$ can only be seen as an approximation to the recombination timescales.

For the \mstar{1.0}{10} galaxy at $z = 3$, as well as for the two galaxies at $z = 0.1$, the evolution of $\Nm{OVI}(R)$ after AGN turn-off is well-described by a sum of two declining exponentials. However, for the \mstar{7.9}{10} galaxy at $z = 3$ (not shown here) a local density and temperature variation at $0.3 < R / R\sub{vir} < 1.3$ causes the decrease of $\Nm{OVI}(R)$ with time to be non-monotonic for the first $3$~Myr after AGN turn-off. After that, $\Nm{OVI}(R)$ decreases monotonically again, with a shape similar to equation (\ref{eq:two_exp}).

\begin{figure*}
\begin{center}
\includegraphics[width=\textwidth]{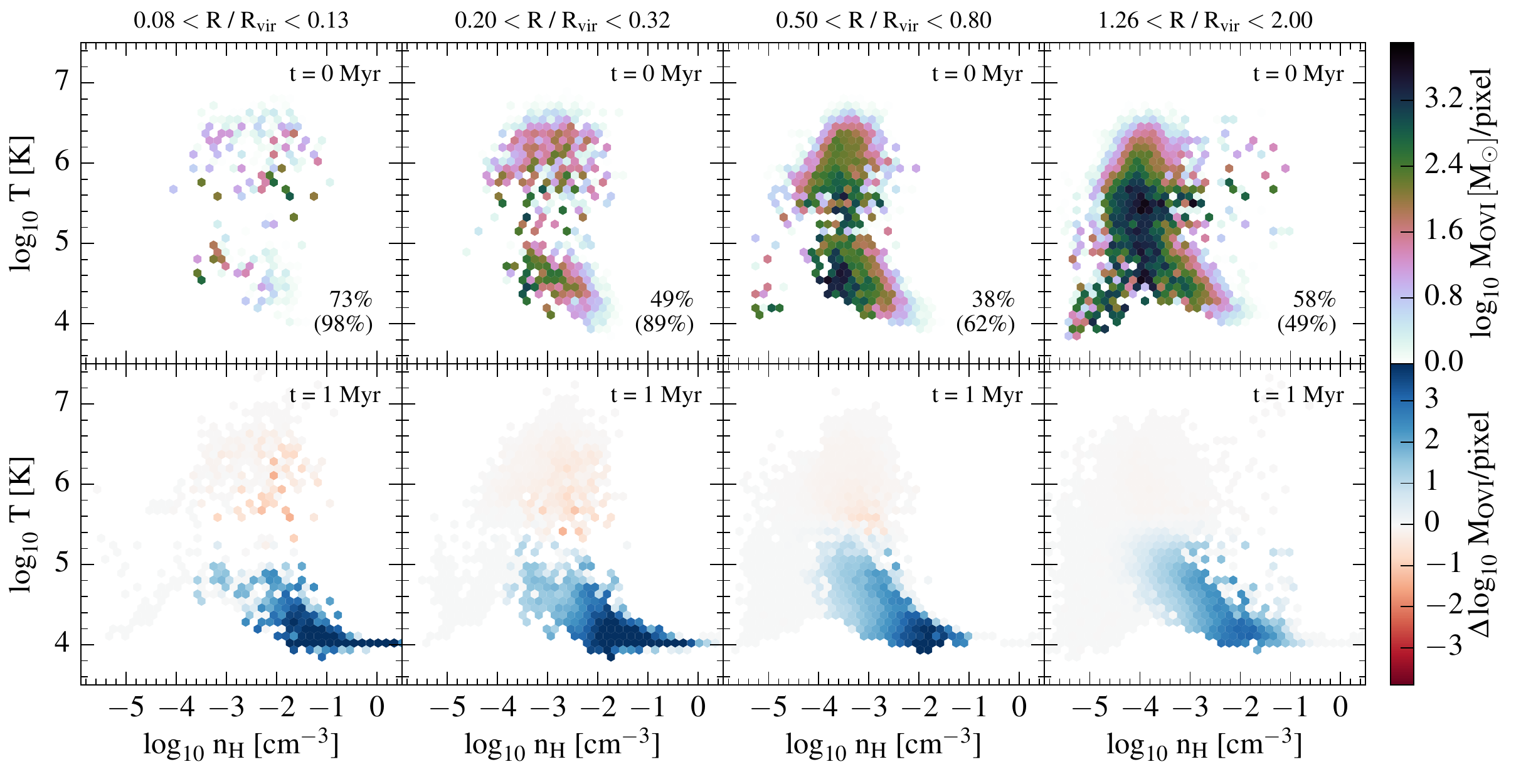}
\end{center}
\caption{The distribution of \OVI mass, $M\sub{OVI}$, in $T - n\sub{H}$ space for different impact parameter intervals for the \mstarsim{10} galaxy at $z=3$. The upper row shows the equilibrium distribution at $t = 0$~Myr, while the lower row shows the difference  in $\log\sub{10} M\sub{OVI}$ per pixel between $t = 1$~Myr (after the AGN has been on for $1$~Myr) and $t = 0$~Myr. In each of the upper panels, the top (bottom) percentage indicates the \OVI mass fraction at $T > 10^5$~K per impact parameter (3D radial distance) bin with boundaries given at the top. The AGN predominantly affects the photoionized gas at $T \lesssim 10^5$~K: the enhancement of \OVI in this temperature regime is what drives the evolution of \N{OVI}.}
\label{fig:example_temp_dens_OVI}
\end{figure*}

In order to investigate at what temperatures and densities the \OVI at different impact parameters arises, and what gas is predominantly affected by the AGN, we plot in Fig.~\ref{fig:example_temp_dens_OVI} the equilibrium \OVI mass distribution (upper row) in $T - n\sub{H}$ space for $4$ $R / R\sub{vir}$ intervals, as well as the difference between the distributions at $t = 1$~Myr and $t = 0$~Myr (lower row). Clearly, at all impact parameters, \OVI occurs in both collisionally ionized ($T \gtrsim 10^5$~K) and photoionized ($T \lesssim 10^5$~K) gas. Contrary to what one might expect, the mass fraction of gas at $T > 10^5$~K (the top percentage indicated in each panel) does not decrease with increasing impact parameter. The reason is that especially the small impact parameter bins include significant quantities of photoionized \OVI residing at large 3D radial distances, because the \OVI profile is relatively flat (see Fig.~\ref{fig:eq_ab_O}). The fraction of $T > 10^5$~K gas per 3D radial distance bin (indicated by the bottom percentage) does, however, decrease with increasing impact parameter, showing that an increasing fraction of the \OVI resides in the photoionized phase\footnote{Although we do not show it here, we find qualitatively similar trends for the other three galaxies.}. This is in qualitative agreement with other theoretical studies of circumgalactic \OVI \citep[e.g.][]{ford_2013,shen_2013}, which generally find \OVI to be mostly collisionally ionized at small galactocentric distances and mostly photoionized at large distances. Furthermore, Fig.~\ref{fig:example_temp_dens_OVI} is also in line with the observations, which show that \OVI can occur in both collisionally ionized and photoionized gas \citep[e.g.][]{carswell_2002,prochaska_2011,savage_2014,turner_2015}.

However, the gas that is most affected by the AGN is the photoionized gas. Apart from a slight decrease of \OVI in the $T \gtrsim 10^{5.3}$~K gas at $R / R\sub{vir} < 0.5$, the main effect is an increase of the \OVI mass at $T \lesssim 10^5$~K. This change in the \OVI abundance at $T \lesssim 10^5$~K is what predominantly drives the evolution of \N{OVI} shown in Fig.~\ref{fig:example_seq_OVI}. Which densities and temperatures dominate \OVI absorption depends on the gas distribution in $T-n\sub{H}$ space and the \OVI ion fraction as a function of $T$ and $n\sub{H}$ \citep[see e.g.][]{oppenheimer_2016}. Due to the additional radiation from the AGN, the \OVI fraction as a function of density in photoionized gas shifts to somewhat higher densities, leading to an increase of the \OVI mass at $n\sub{H} = 10^{-4} - 10^{-1}$~cm$^{-3}$. Note that this density range in which \OVI is enhanced is roughly the same at all impact parameters, even though the typical density of CGM gas decreases with increasing impact parameter (as, for example, seen in the upper panels). The corresponding re-equilibration timescale of \N{OVI} after AGN turn-off is therefore also expected to be roughly independent of impact parameter. This is consistent with Fig.~\ref{fig:example_seq_OVI}, where at all $R / R\sub{vir} > 0.2$ $\Nm{OVI}(R)$ reaches $37 \%$ of its peak value (i.e. approximately the e-folding timescale) $\approx 4 - 5$~Myr after the AGN turns off (correcting for the light-travel time delay). At $0.08 < R / R\sub{vir} < 0.2$, this timescale is slightly shorter, $\approx 2 - 3$~Myr, mainly due to a deficit of low-density gas.


\subsection{Dependence on impact parameter}
\label{sec:rad_dep}

In this and the next section we investigate the strength of the AGN fossil effect, quantified by the deviation in the average \OVI column density and covering fraction from the respective equilibrium values, in the CGM of the four galaxies shown in Fig.~\ref{fig:galaxies}. For all galaxies, we adopt the same AGN lifetime and duty cycle fraction as in the previous section: $\left\lbrace \tagnm = 10^6 \, \mathrm{yr}, \fdutym = 10 \% \right\rbrace$. For the Eddington ratio, we take $\eddratiom = 1.0$ at $z = 3$ and $\eddratiom = 0.1$ at $z = 0.1$.

Fig.~\ref{fig:mass_dep_OVI} shows \avN{OVI} (solid lines; left panels), as defined in equation~(\ref{eq:av_N}), and \avfcov{OVI} (solid lines; right panels), as defined in equation~(\ref{eq:av_f_cov}), as a function of normalized impact parameter for the \mstarsim{10} (blue) and \mstarsim{11} (red) galaxies at $z = 0.1$ (upper panel) and $z = 3$ (lower panel). The column density and covering fraction profiles in equilibrium are indicated by dashed lines. For all four galaxies, the deviation in \avN{OVI} and \avfcov{OVI} from equilibrium decreases with increasing impact parameter. This is mainly because the flux of ionizing photons from the AGN decreases as $R\up{-2}$, but also because the column densities of the \OI{ }- \OV oxygen ions decrease with increasing impact parameter (see Fig.~\ref{fig:eq_ab_O}). This causes a larger initial offset in $\Nm{OVI}(R)$ -- and related to this, a larger initial offset in $\fcovm{OVI}(R)$ -- at small $R / R\sub{vir}$, while the re-equilibration timescale on which $\Nm{OVI}(R)$ and $\fcovm{OVI}(R)$ decrease after AGN turn-off is roughly the same at all $R / R\sub{vir}$. In general, the fact that there is a significant offset in \avN{OVI} and \avfcov{OVI} over the whole $R / R\sub{vir}$ range for all four galaxies, indicates that the recombination timescales in the CGM are sufficiently long to establish AGN proximity zone fossils out to at least twice the virial radius from \mstarsim{10} and \mstarsim{11} galaxies at both $z = 0.1$ and $z = 3$.


\subsection{Dependence on galaxy mass and redshift}
\label{sec:mass_dep}

\begin{figure*}
\begin{center}
\includegraphics[width=\textwidth]{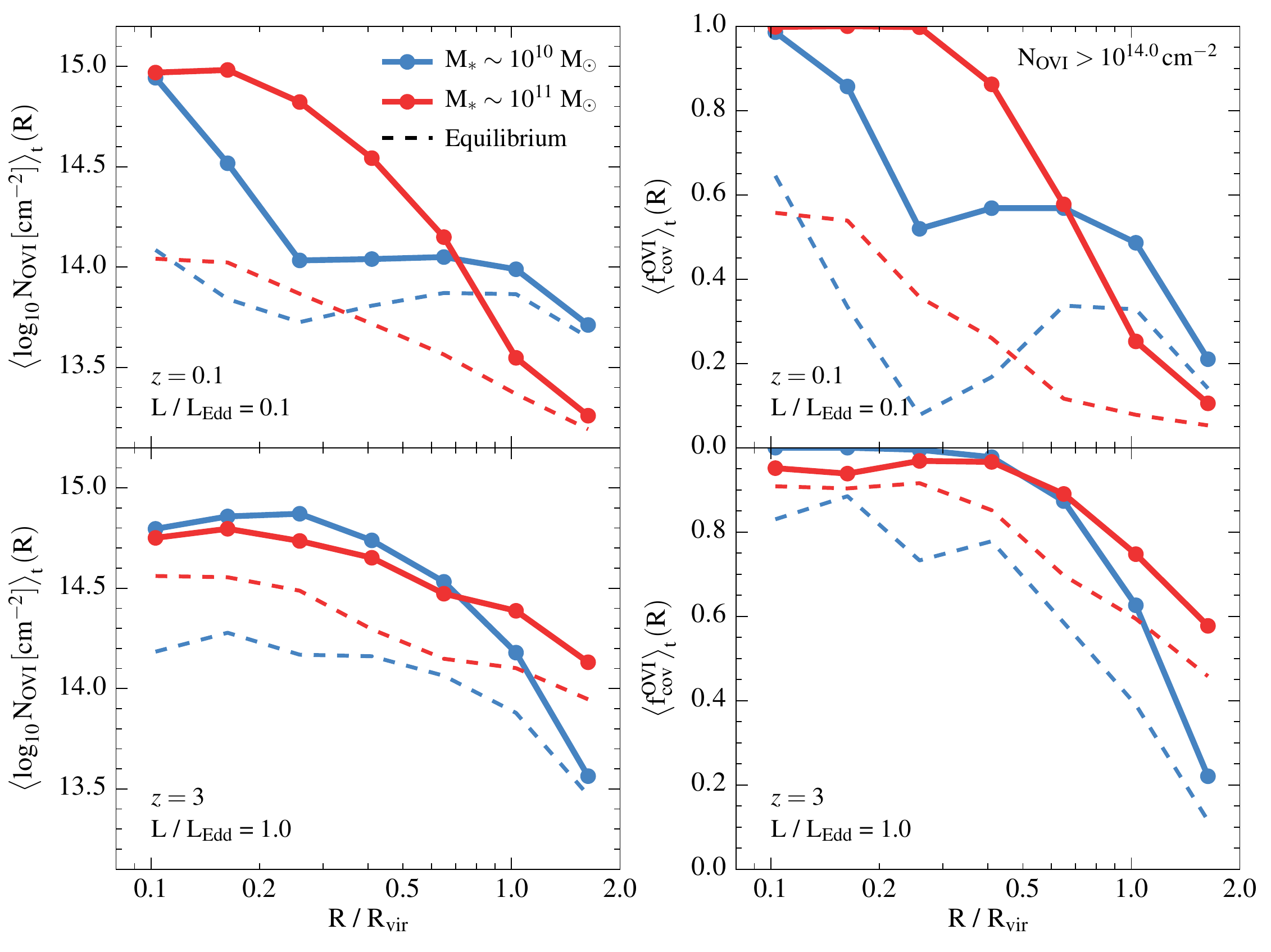}
\end{center}
\caption{The strength of the AGN fossil effect as a function of impact parameter, normalized by the virial radius, for the \mstarsim{10} (blue) and \mstarsim{11} (red) galaxies at $z = 0.1$ (upper panels) and $z = 3$ (lower panels). The adopted AGN parameters are $\tagnm = 10^6$~yr and $\fdutym = 10 \%$, with $\eddratiom = 0.1$ at $z = 0.1$ and $\eddratiom = 1.0$ at $z = 3$. The solid curves show the average \OVI column density, \avN{OVI} (equation~\ref{eq:av_N}; left panels), and average covering fraction of $\Nm{OVI} > 10^{14.0}$ cm$^{-2}$ gas, \avfcov{OVI} (equation~\ref{eq:av_f_cov}; right panels), in $7$ logarithmic impact parameter bins of size $0.2$~dex. The dashed curves show the corresponding profiles in equilibrium. At $z = 3$, the fossil effect at $R / R\sub{vir} \lesssim 0.8$ around the \mstarsim{10} galaxy is significantly larger than around the \mstarsim{11} galaxy, despite the $\approx 2$ times lower AGN flux that the gas at a fixed $R / R\sub{vir}$ receives. This is due to the larger abundance of \OI{ }- \OV oxygen ions that can be ionized to \OVI. At $z = 0.1$, the fossil effect is largest around the \mstarsim{11} galaxy over the whole impact parameter range. Even though the gas at a fixed $R / R\sub{vir}$ at $z = 0.1$ receives $60 - 80$ times lower AGN flux than the gas around a similarly massive galaxy at $z = 3$, the fossil effect at $z = 0.1$ is often larger, owing to the $\sim 10$ times longer \OVI re-equilibration timescale.}
\label{fig:mass_dep_OVI}
\end{figure*}

As we already discussed in Section~\ref{sec:galaxies}, the column density of \OVI and the relative abundances of the different oxygen ions (i.e. the overall ionization state of the gas) are sensitive to a number of factors that are related to the mass of the galaxy -- like the halo virial temperature and the galaxy metallicity. Also, more massive galaxies host more massive BHs \citep[e.g.][]{merritt+ferrarese_2001}, suggesting that the AGN are also more luminous during their active phase. Hence, we expect the effect of a fluctuating AGN on the CGM to be dependent on the mass of the galaxy. Furthermore, since AGN are generally observed to be more luminous at higher redshift \citep[e.g.][]{kollmeier_2006}, and the density of the Universe increases with increasing redshift, the effect is not necessarily quantitatively similar for galaxies of a similar stellar mass at different redshifts.

To investigate the dependence of the AGN fossil effect on galaxy stellar mass, we start by comparing the \mstarsim{10} and \mstarsim{11} galaxies at $z = 3$ (lower panels of Fig.~\ref{fig:mass_dep_OVI}). For a fluctuating AGN in the \mstarsim{11} galaxy, \avN{OVI} and \avfcov{OVI} are generally $\approx 0.2 - 0.4$~dex and $\approx 0.04 - 0.2$, respectively, higher than in the equilibrium case, and the offsets change only mildly with impact parameter. For the \mstarsim{10} galaxy, however, the significance of the fossil effect changes more rapidly, causing the effect to be somewhat smaller than for the \mstarsim{11} galaxy at large impact parameters ($R / R\sub{vir} \gtrsim 1.3$), but significantly larger at small impact parameters ($R / R\sub{vir} \lesssim 0.8$). For $R / R\sub{vir} < 0.5$, the offsets in \avN{OVI} and \avfcov{OVI} with respect to equilibrium increase to $\gtrsim 0.5$~dex and $\gtrsim 0.2$, respectively.

Similar to the dependence on impact parameter, the dependence of the fossil effect on galaxy stellar mass can be explained by considering the difference in the ionizing flux, the re-equilibration timescale and the abundance of low-state ions. For an AGN with a fixed Eddington ratio, the flux at a fixed $R / R\sub{vir}$ scales with the stellar mass as $\propto M\sub{\ast}\up{1/3}$, since $L\sub{Edd} \propto M\sub{BH} \propto M\sub{\ast}$ and $R\sub{vir} \propto M\sub{vir}\up{1/3} \propto M\sub{\ast}\up{1/3}$ approximately. The gas around the \mstarsim{11} galaxy therefore receives a $\approx 2$ times higher AGN flux than the gas at the same $R / R\sub{vir}$ receives from the AGN in the \mstarsim{10} galaxy. The re-equilibration timescale of \OVI is similar for both galaxies, since the \OVI (enhancement) occurs at similar densities at a fixed fraction of the virial radius. Hence, the fact that we find a larger fossil effect at $R / R\sub{vir} \lesssim 0.8$ around the \mstarsim{10} galaxy must be due to a larger abundance of low-state oxygen ions (see Fig.~\ref{fig:eq_ab_O}). This causes a larger initial boost in $\log\sub{10} \Nm{OVI}(R)$ than for the \mstarsim{11} galaxy, despite the fact that the flux from the AGN is lower. At large impact parameters ($R / R\sub{vir} \gtrsim 1.3$), however, the fossil effect is larger for the \mstarsim{11} galaxy, which can be attributed to a combination of higher AGN flux and the larger abundance of low-state ions in this $R / R\sub{vir}$ regime.

The opposite trend with stellar mass is seen at $z = 0.1$ (upper panels): the fossil effect around the \mstarsim{11} galaxy is much larger than around the \mstarsim{10} galaxy (except for the innermost $R / R\sub{vir}$ bin), and increases strongly with decreasing impact parameter. The difference is particularly evident at $0.2 \lesssim R / R\sub{vir} \lesssim 0.8$, where the offsets of \avN{OVI} and \avfcov{OVI} from equilibrium are $\approx 0.6 - 1.0$~dex and $\approx 0.5 - 0.6$, respectively, in the high-mass case and $\approx 0.2 - 0.3$~dex and $\approx 0.2 - 0.4$, respectively, in the low-mass case.

When comparing galaxies with a similar stellar mass at different redshifts, the re-equilibration timescale does play an important role. Since the virial radii of the two galaxies at $z = 0.1$ are $2.5 - 2.8$ times larger than those of the similarly massive galaxies at $z = 3$, and the AGN implemented at $z = 0.1$ have a $10$ times lower Eddington ratio, gas at a fixed $R / R\sub{vir}$ receives a $60 - 80$ times lower AGN flux at $z = 0.1$ than at $z = 3$. In combination with the lower abundance of low-state oxygen ions, it may seem surprising that we see a fossil effect at $z = 0.1$ at all. The reason is the significantly longer recombination timescales in the CGM: by comparing figures similar to Fig.~\ref{fig:example_temp_dens_OVI} for all four galaxies, we find that the AGN-induced \OVI enhancement (in a fixed $R / R\sub{vir}$ interval) occurs at $\sim 10$ times lower densities for the galaxies at $z=0.1$ than for those at $z=3$. Hence, the expected re-equilibration timescale of \OVI is $\sim 10$ times longer at $z=0.1$ than at $z=3$. For the $z = 0.1$ galaxy with \mstarsim{10}, the offsets of \avN{OVI} and \avfcov{OVI} from equilibrium at $R / R\sub{vir} \gtrsim 0.2$ are still smaller than for its high-redshift equivalent. However, the offsets are large enough to substantially enhance \OVI in low-redshift CGM observations, as is explored by \citet{oppenheimer_2017}. For the $z = 0.1$ galaxy with \mstarsim{11}, the fossil effect at impact parameters of $R / R\sub{vir} \lesssim 0.8$ does become much larger than at high redshift, owing to the significantly prolonged recombination phase after AGN turn-off.

To quantify the probability of observing a significant AGN fossil effect, we calculate the fraction of the time in between two AGN-on phases for which $\log\sub{10} \Nm{OVI}$ is offset from the equilibrium value by $> 0.1$~dex. This can be interpreted as the probability that an observation of circumgalactic \OVI is significantly affected by AGN fossil effects, even though the galaxy would not be identified as an active AGN host. For the \mstarsim{11} galaxy at $z = 3$, this probability varies between $\approx 0.5$ and $\approx 1.0$ over the whole impact parameter range, while for the \mstarsim{10} galaxy at $z = 3$, it is $\approx 1.0$ for $R / R\sub{vir} < 1.3$ and drops steeply at higher $R / R\sub{vir}$ (although still being $\approx 0.4$ for $1.3 < R / R\sub{vir} < 2.0$). Hence, the \OVI around the \mstarsim{10} galaxy out to $R / R\sub{vir} = 1.3$ is constantly kept out of equilibrium, even though the AGN duty cycle fraction is only $10 \%$. At $z = 0.1$, the probability of observing a significant fossil effect changes even more abruptly with impact parameter: while it is $\approx 1.0$ for $R / R\sub{vir} < 1.3$, the probability is zero at larger $R / R\sub{vir}$, independent of galaxy mass. Hence, we expect that both at $z = 0.1$ and at $z = 3$ galaxies with $M\sub{\ast} = 10^{10 - 11} \, \mathrm{M}\sub{\odot}$ are significantly affected by AGN fossil effects: AGN with the luminosity, lifetime and duty cycle fraction adopted here, keep circumgalactic \OVI continuously out of equilibrium up to impact parameters that extend beyond the virial radius ($R / R\sub{vir} < 1.3$).


\subsection{Dependence on AGN properties}
\label{sec:agn_dep}

\subsubsection{AGN luminosity}
\label{sec:lum_dep}

\begin{figure*}
\begin{center}
\includegraphics[width=\textwidth]{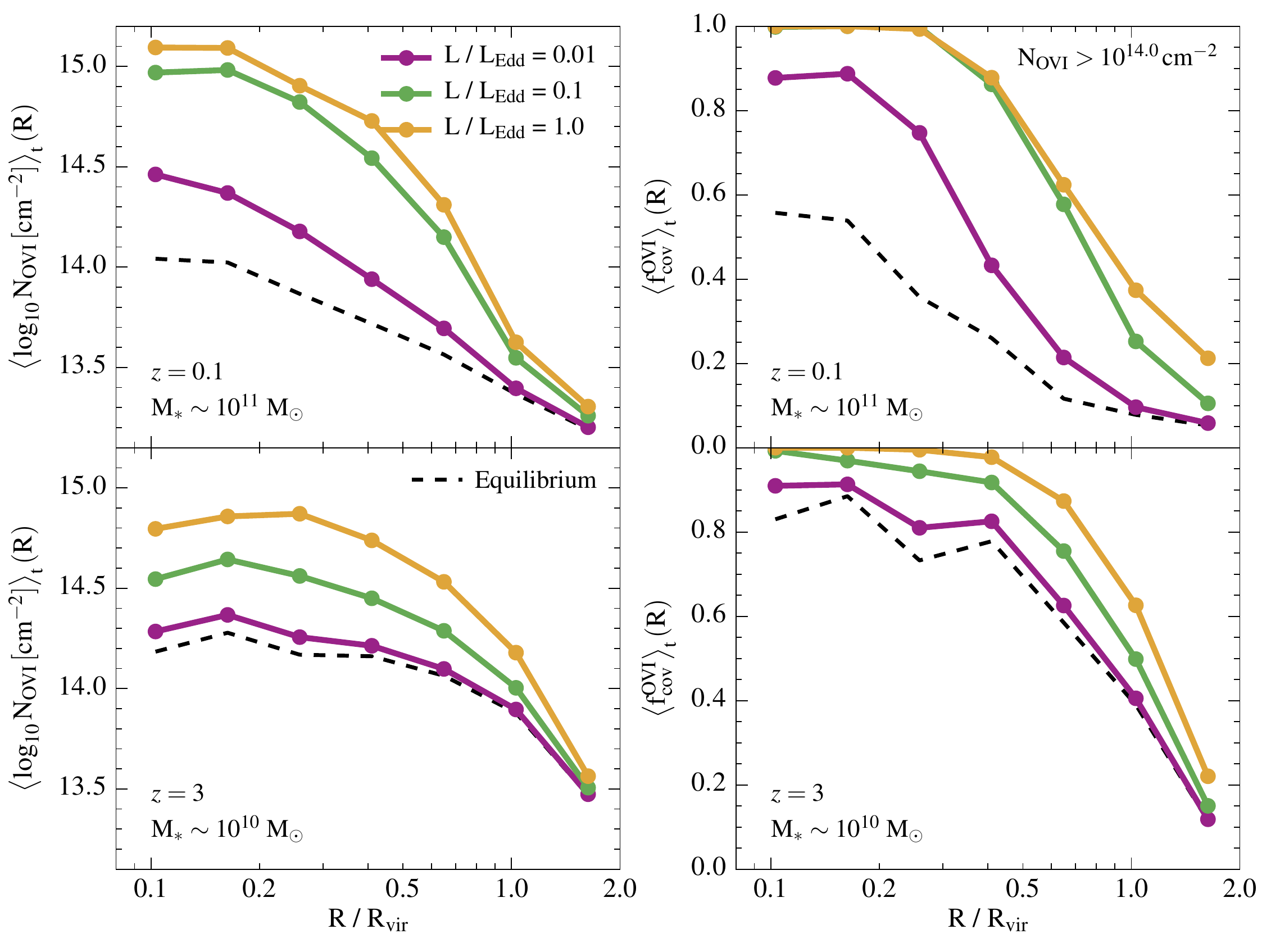}
\end{center}
\caption{The impact of the AGN Eddington ratio on the fossil effect for the \mstarsim{11} galaxy at $z = 0.1$ (upper panels) and the \mstarsim{10} galaxy at $z = 3$ (lower panels). The plotted quantities, including the equilibrium profiles for the respective galaxies (black, dashed lines), are the same as in Fig.~\ref{fig:mass_dep_OVI}, but now we compare different Eddington ratios: $\eddratiom = 0.01$ (purple), $\eddratiom = 0.1$ (green) and $\eddratiom = 1.0$ (yellow). We still adopt $\tagnm = 10^6$~yr and $\fdutym = 10 \%$. For a sufficiently high AGN luminosity, as in the case of the \mstarsim{11} galaxy at $z = 0.1$, the increase in \avN{OVI} and \avfcov{OVI} with respect to equilibrium depends on the adopted luminosity in a non-linear way. This is related to the increase in the photoionization rate from \OVI to higher states as well as from \OV to \OVI.}
\label{fig:lum_dep_OVI}
\end{figure*}

Observationally, AGN are found to exhibit a wide variety of Eddington ratios, which are generally smaller and span a larger range of values at low redshift than at high redshift \citep[e.g.][]{kollmeier_2006,kauffmann+heckman_2009}. We therefore explore the dependence of the AGN fossil effect on the Eddington ratio, mainly focusing on $z = 0.1$. We adopt the same \tagn and \fduty as in the previous sections.

The upper panels of Fig.~\ref{fig:lum_dep_OVI} compare our fiducial $\eddratiom = 0.1$ (green) to a $10$ times lower (purple) and a $10$ times higher (yellow) Eddington ratio for the \mstarsim{11} galaxy at $z=0.1$. The ratio $\eddratiom = 0.1$ corresponds to a (bolometric) luminosity of $L = 1.3 \times 10^{45}$~erg/s, or $L\sub{0.5-2~\mathrm{keV}} = 2.6 \times 10^{43}$~erg/s in the soft X-ray band ($0.5 - 2$~keV) and $L\sub{2-10~\mathrm{keV}} = 3.5 \times 10^{43}$~erg/s in the hard X-ray band ($2 - 10$~keV), which we calculate by integrating the AGN spectrum over the appropriate energy interval. These $L\sub{0.5-2~\mathrm{keV}}$ and $L\sub{2-10~\mathrm{keV}}$ correspond to AGN strengths that are commonly observed at $z\approx 0.1$, where these luminosities are close to the knee of the X-ray AGN luminosity function \citep[e.g.][]{ueda_2003,hasinger_2005,aird_2010}\footnote{The knee shifts to approximately an order of magnitude higher luminosity from $z = 0$ to $z = 3$, justifying our choice of adopting a $10$ times higher fiducial Eddington ratio at $z = 3$ than at $z = 0.1$.}. The value $\eddratiom = 1.0$ then represents a more extreme case, while $\eddratiom = 0.01$ corresponds to a more moderate AGN.

Over the whole impact parameter range, the offsets of \avN{OVI} and \avfcov{OVI} for $\eddratiom = 0.01$ are significantly smaller than those for $\eddratiom = 0.1$. Within $R / R\sub{vir} = 0.8$, the offsets are $\approx 0.1 - 0.4$~dex and $\approx 0.1 - 0.4$, respectively, for $\eddratiom = 0.01$, compared to $\approx 0.6 - 1.0$~dex and $\approx 0.4 - 0.6$ for $\eddratiom = 0.1$. The $10$ times lower Eddington ratio also causes the fossil effect to be observable out to smaller impact parameters: \OVI is observed to be continuously out of equilibrium (i.e. a probability of $1.0$) out to $0.8 R / R\sub{vir}$ instead of $1.3 R / R\sub{vir}$. While the $\eddratiom = 0.01$ AGN therefore affects a $\approx 4$ times smaller volume in the CGM of individual galaxies, they are $\approx 10$ times more common than the $\eddratiom = 0.1$ AGN according to the luminosity function from \citet[][comparing the occurence of $L\sub{2-10~\mathrm{keV}} = 3.5 \times 10^{42}$~erg/s to $L\sub{2-10~\mathrm{keV}} = 3.5 \times 10^{43}$~erg/s]{aird_2010}. Hence, the total volume in the Universe that these low-luminosity AGN are expected to affect is larger.

In the high-Eddington ratio ($\eddratiom = 1.0$) case, the offsets of \avN{OVI} and \avfcov{OVI} from equilibrium are larger than in the $\eddratiom = 0.1$ case. However, the difference is not as significant as between $\eddratiom = 0.1$ and $\eddratiom = 0.01$. This is related to the fact that adopting a $10$ times higher AGN luminosity does not only increase the photoionization rate from \OV to \OVI, but also increases the rate from \OVI to \OVII and to higher oxygen states. Hence, the increase in \avN{OVI} and \fcov{OVI} is not linearly dependent on the AGN luminosity. These quantities can even show a deficit with respect to equilibrium for extremely high AGN luminosities, when most of the oxygen ions are ionized from \OI{ }- \OVI to higher states. We see a similar effect if we consider metal ion species with a lower ionization energy than \OVI (see Section~\ref{sec:other_metals}).

Despite the limited increase in the fossil effect from $\eddratiom = 0.1$ to $\eddratiom = 1.0$, the high-Eddington ratio AGN establishes a continuously observable fossil effect out to the largest impact parameters that we consider here. While this affects a $\approx 4$ times larger volume around individual galaxies than in the $\eddratiom = 0.1$ case, AGN with $L\sub{2-10~\mathrm{keV}} = 3.5 \times 10^{44}$~erg/s are $\approx 20 - 30$ times less common than AGN with $L\sub{2-10~\mathrm{keV}} = 3.5 \times 10^{43}$~erg/s \citep{aird_2010}, thereby affecting a smaller total volume in the Universe.

The lower panels of Fig.~\ref{fig:lum_dep_OVI} show the dependence of the high-redshift fossil effect on the AGN Eddington ratio, where we focus on the \mstarsim{10} galaxy at $z = 3$. Increasing the Eddington ratio from $\eddratiom = 0.01$ to $\eddratiom = 0.1$ has a similar effect on \avN{OVI} and \avfcov{OVI} as increasing the ratio from $\eddratiom = 0.1$ to $\eddratiom = 1.0$. The increasing photon flux primarily enhances the ionization from low oxygen states to \OVI: there is no `saturation' of the \OVI enhancement like at $z = 0.1$. Furthermore, the faint-end slope of the X-ray luminosity function at $z = 3$ is flatter than at low redshift, making AGN with a $10$ times lower \eddratio only $\approx 2.5 - 4$ more common (\citealp{aird_2010}; comparing the occurrence of the corresponding $L\sub{2-10~\mathrm{keV}}$ values, which are all well below the knee of the luminosity function at $z = 3$). Hence, in contrast with low redshift, the more luminous AGN are expected to affect a larger volume in the Universe than the low-luminosity AGN, as the AGN with $\eddratiom = 1.0$ ($\eddratiom = 0.1$) affects a $\approx 200$ ($\approx 40$) times lager volume around individual galaxies than the AGN with $\eddratiom = 0.01$.


\subsubsection{AGN lifetime and duty cycle}
\label{sec:cycle_dep}

\begin{figure*}
\begin{center}
\includegraphics[width=\textwidth]{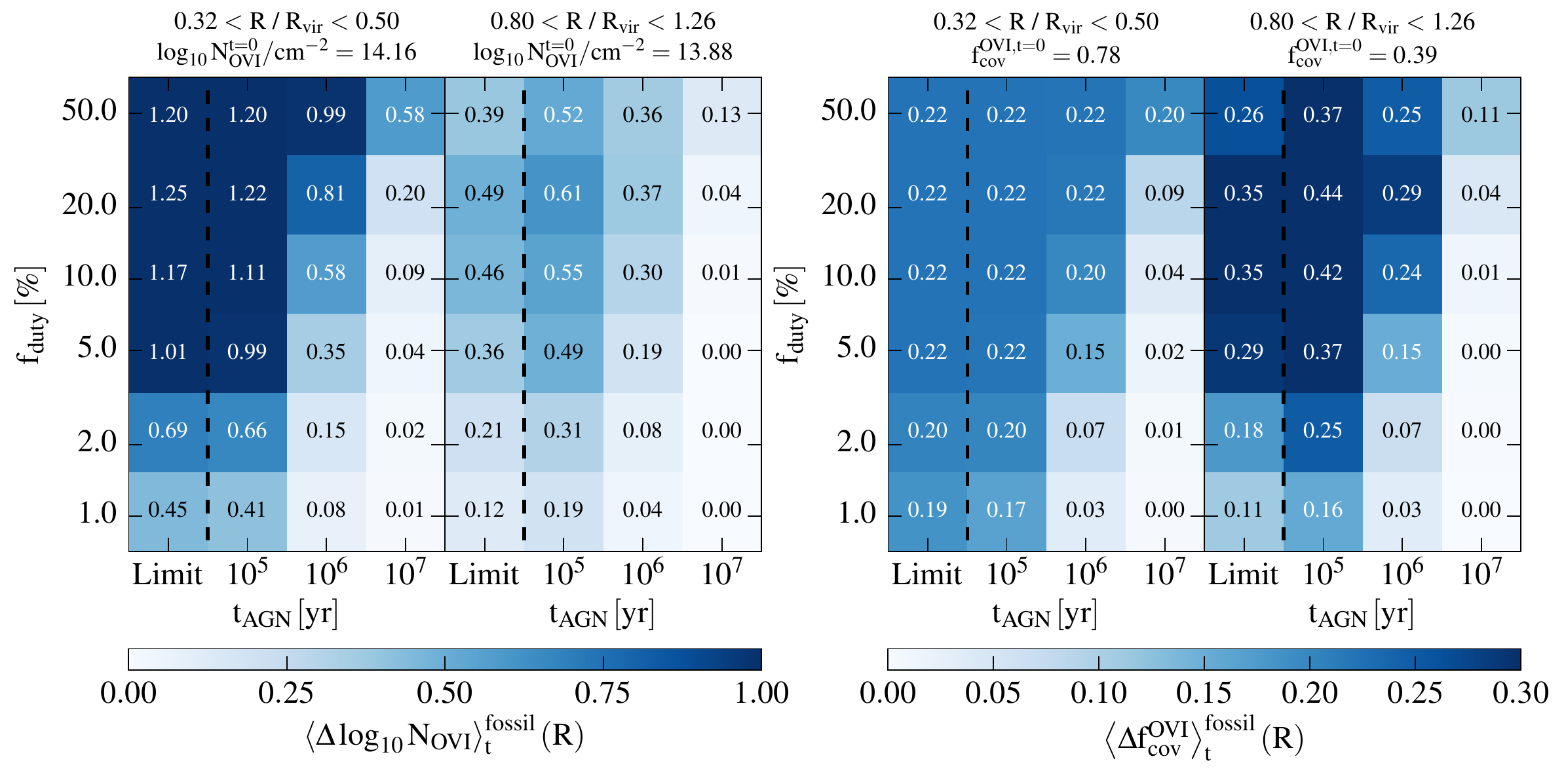}
\end{center}
\caption{The impact of varying the AGN lifetime and duty cycle fraction on the strength of the AGN fossil effect for the \mstarsim{10} galaxy at $z = 3$ (with an AGN with $\eddratiom = 1.0$). The left-hand and right-hand panels show the average out-of-equilibrium offsets of the \OVI column density and covering fraction (equation~\ref{eq:delta_av_N}), repectively. We focus on two impact parameter intervals. For each \tagn and \fduty, the offsets are shown as both a colour-coding and a number, while the corresponding equilibrium values are indicated at the top. The fossil effect tends to be stronger for shorter AGN lifetimes and larger duty cycle fractions, as the time in between subsequent AGN-on phases is shorter. In the limit of very short AGN lifetimes, the effect of a fluctuating AGN on the \OVI column density and covering fraction converges to the effect of a continuously radiating AGN with a luminosity equal to $(\fdutym / 100 \%)$ times the original luminosity (leftmost column in each panel: `Limit').}
\label{fig:av_map_OVI}
\end{figure*}

The wide range of observational estimates of \tagn and \fduty presented in the literature shows that either AGN exhibit a large variety of lifetimes and duty cycle fractions, or that these parameter values are highly uncertain (as they are mostly based on indirect constraints). We therefore explore the impact of \tagn and \fduty on the AGN fossil effect. We focus on the \mstarsim{10} galaxy at $z = 3$, but for high stellar mass or low redshift the conclusions are qualitatively similar.

In Fig.~\ref{fig:av_map_OVI} we show
\begin{equation}\label{eq:delta_av_N}
\avNfossilm{OVI} \equiv \avNm{OVI} - \log\sub{10} \Nm{OVI}\up{t=0}(R)
\end{equation}
in the left two panels and \avfcovfossil{OVI}, defined in a similar way, in the right two panels, for different choices of \tagn and \fduty (the extra column called `Limit' will be discussed below). We focus here on two impact parameter intervals. Both quantities show that the AGN fossil effect tends to become stronger towards smaller AGN lifetimes and larger duty cycle fractions. This is related to the fact that the AGN-off time (i.e. the time in between two subsequent AGN-on phases) decreases with decreasing \tagn and increasing \fduty. For example, while $t\sub{off} = 90$~Myr for $\tagnm = 10^7$~yr and $\fdutym = 10 \%$, $t\sub{off} = 0.9$~Myr for $\tagnm = 10^5$~yr and $\fdutym = 10 \%$. This is a significant difference considering the typical re-equilibration timescale of \OVI (i.e. the $\Nm{OVI}(R)$ enhancement returns to $37 \%$ of its peak value in $\approx 2 - 5$~Myr). The small $t\sub{off}$ leads to average enhancements in the column density and covering fraction up to $\avNfossilm{OVI} = 1.22$~dex and $\avfcovfossilm{OVI} = 0.22$ (i.e. reaching $\fcovm{OVI} = 1.0$) for $0.3 < R / R\sub{vir} < 0.5$, and up to $\avNfossilm{OVI} = 0.61$~dex and $\avfcovfossilm{OVI} = 0.44$ for $0.8 < R / R\sub{vir} < 1.3$.

In general, an AGN proximity zone fossil is created most efficiently if the AGN lifetime is comparable to or longer than the ionization timescale from \OV to \OVI (e.g. $\sim 10^5$~yr at $R / R\sub{vir} \approx 0.8$), so as to ionize significant quantities of oxygen to \OVI, and if the AGN-off time is similar or shorter than the re-equilibration timescale of \OVI. For $\tagnm = 10^5$~yr this is roughly the case for $\fdutym = 2 \%$ (where $t\sub{off} = 4.9$~Myr), for $\tagnm = 10^6$~yr at $\fdutym = 20 \%$ (where $t\sub{off} = 4.0$~Myr) and for $\tagnm = 10^7$~yr at $\fdutym \gtrsim 50 \%$ (where $t\sub{off} \lesssim 10$~Myr). This approximately corresponds to the regime in Fig.~\ref{fig:av_map_OVI} where we find a significant AGN fossil effect.

\begin{figure}
\begin{center}
\includegraphics[width=\columnwidth]{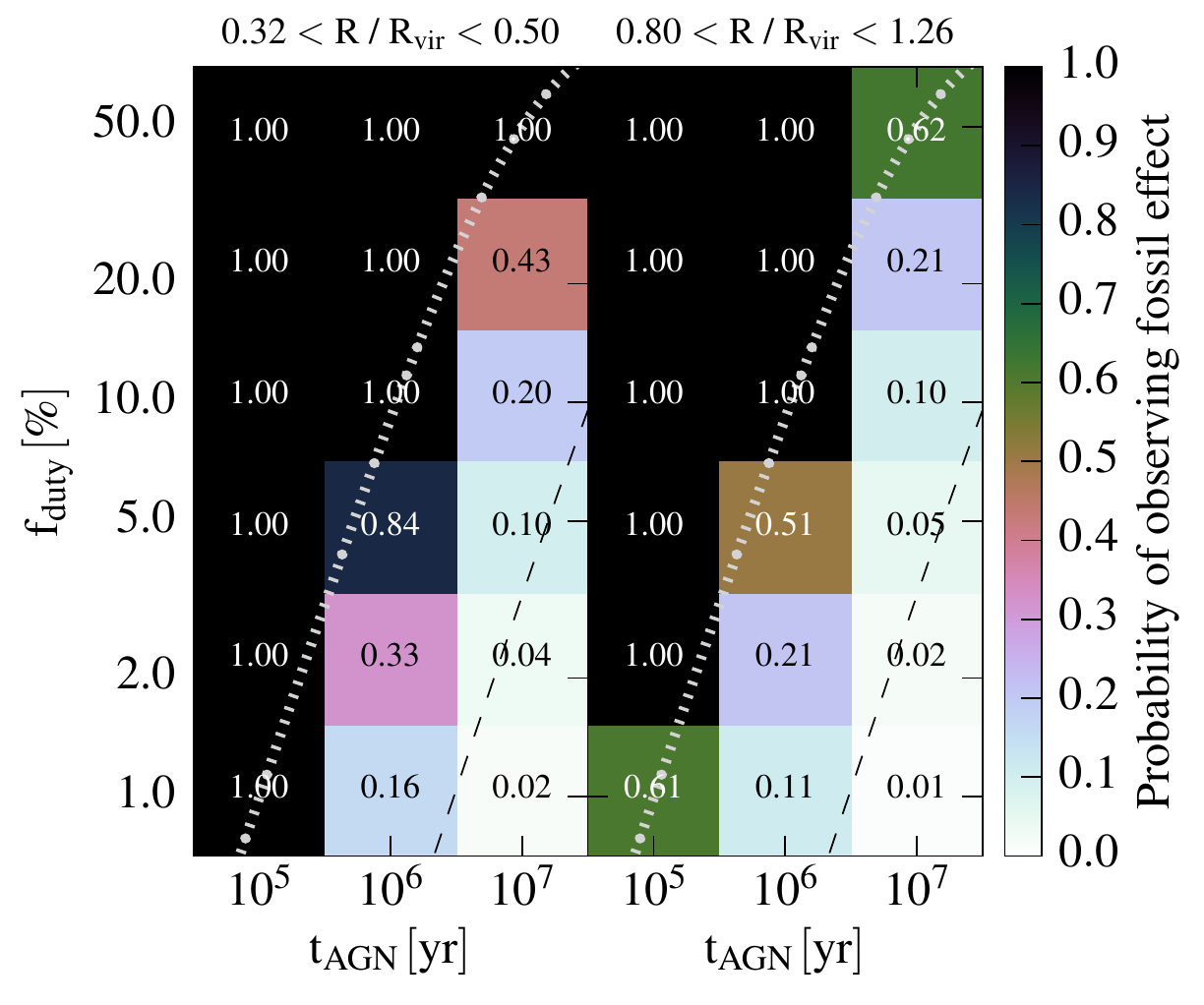}
\end{center}
\caption{The impact of varying the AGN lifetime and duty cycle fraction on the probability of observing a significant AGN fossil effect (i.e. $\log\sub{10} \Nm{OVI}$ offset from equilibrium by $> 0.1$~dex during the AGN-off time) for the \mstarsim{10} galaxy at $z = 3$ with a $\eddratiom = 1.0$ AGN. If the time in between subsequent AGN-on phases is $t\sub{off} \lesssim 10$~Myr (to the left of the grey, dotted lines), the \OVI is continuously kept in an out-of-equilibrium state, and the \OVI enhancement accumulates over multiple cycles. For $t\sub{off} \gtrsim 300$~Myr (to the right of the black, dashed line), the time in between subsequent AGN-on phases is too long to see a significant fossil effect for a significant fraction of the time.}
\label{fig:av_map_frac_OVI}
\end{figure}

\begin{figure}
\begin{center}
\includegraphics[width=\columnwidth]{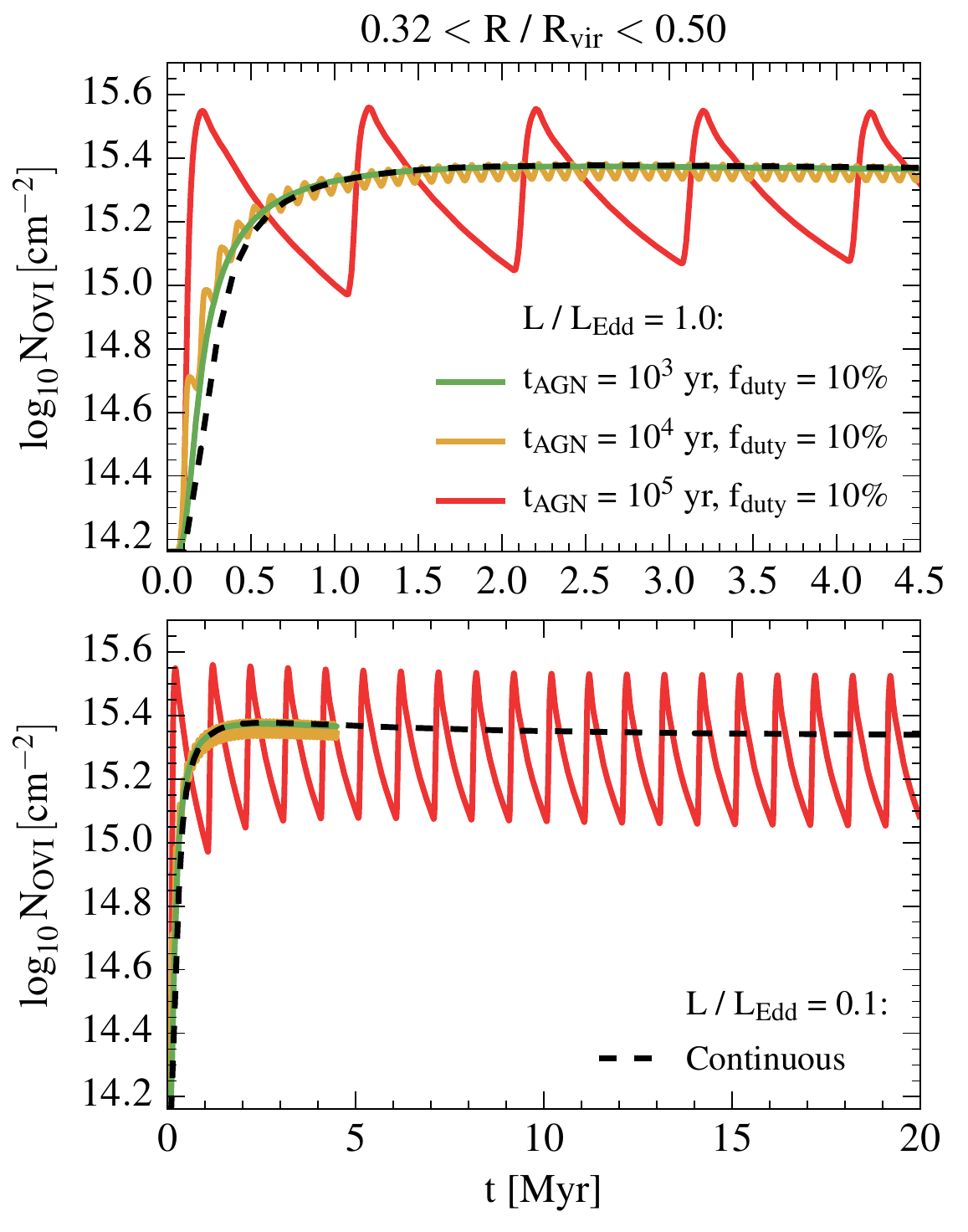}
\end{center}
\caption{The \OVI column density at $0.3 < R / R\sub{vir} < 0.5$ from the \mstarsim{10} galaxy at $z = 3$ as a function of time, for different values of the AGN lifetime: $\tagnm = 10^3$~yr (green, solid line), $\tagnm = 10^4$~yr (yellow, solid line) and $\tagnm = 10^5$~yr (red, solid line). The AGN has $\eddratiom = 1.0$ and is fluctuating with a constant duty cycle fraction of $\fdutym = 10 \%$. The upper panel zooms in on the $0 < t < 4.5$~Myr range of the lower panel. Since the time in between AGN-on phases for these combinations of \tagn and \fduty is sufficiently short, the \OVI enhancement accumulates over multiple cycles, until a new equilibrium (cycle average) has been reached. In the limit of small \tagn, the net evolutionary trend converges to a trend corresponding to an AGN continuously radiating at $\eddratiom = \fdutym / 100 \% = 0.1$ (black, dashed line).}
\label{fig:asymp_OVI}
\end{figure}

The probability of observing a significant fossil effect while the AGN is inactive, which is shown as a function of \tagn and \fduty in Fig.~\ref{fig:av_map_frac_OVI}, is close to or equal to $1.0$ if $t\sub{off} \lesssim 10$~Myr (to the left of the grey, dotted line). This means that throughout the whole AGN-off time, \N{OVI} is offset from equilibrium by at least $0.1$~dex, indicating that the \OVI is constantly kept in an overionized state. This limit of $t\sub{off} \approx 10$~Myr is roughly independent of impact parameter. The fact that \N{OVI} is continuously kept out of equilibrium also causes the enhancement of \N{OVI} to build up over many cycles. This is illustrated for an AGN with $\tagnm = 10^5$~yr and $\fdutym = 10 \%$ (red, solid lines) in Fig.\ref{fig:asymp_OVI}, for the same galaxy and AGN set-up as in Figs.~\ref{fig:av_map_OVI} and \ref{fig:av_map_frac_OVI}, and for $0.3 < R / R\sub{vir} < 0.5$. The net evolutionary trend is an increase of \N{OVI} over a timescale of a few megayears. Eventually, the increase flattens off and \OVI reaches a new equilibrium, where the combined photoionization from the HM01 background and the AGN (in addition to collisional ionization) balances the recombinations per cycle. Similar asymptotic behaviour is shown by the \OVI covering fraction. As we mentioned in Section~\ref{sec:quant_agn_fossil}, we calculate \avN{OVI} and \avfcov{OVI} only after such a new ionization balance has been reached (for this combination of \tagn and \fduty and other combinations in Figs.~\ref{fig:av_map_OVI} and \ref{fig:av_map_frac_OVI} to which this applies).

If we now consider AGN with $\tagnm = 10^4$~yr (yellow, solid line) and $\tagnm = 10^3$~yr (green, solid line), while keeping the duty cycle fraction the same at $\fdutym = 10\%$, we find that in both cases the net evolution is the same as for $\tagnm = 10^5$~yr. It just takes more cycles, with smaller individual fluctuations, when \tagn is shorter\footnote{Note that in this way, despite the fact that the AGN lifetime is shorter than the ionization timescale of \OV to \OVI, the fluctuating AGN still give rise to a significant fossil effect.}. This suggests that, as could have been expected, in the limit of $\tagnm \to 0$, the behaviour of \N{OVI} converges to the result obtained for an AGN radiating continuously at $10 \%$ of the original flux. Indeed, a continuously radiating AGN with $\eddratiom = 0.1$ (black, dashed line) traces the net \N{OVI} evolution of the fluctuating AGN with $\eddratiom = 1.0$ and $\fdutym = 10 \%$\footnote{Note that after the initial (net) increase in \N{OVI} for $0 < t < 2$~Myr, \N{OVI} first decreases slightly -- corresponding to the ionization of \OVI to higher states -- before a new equilibrium is established. This is seen in the trends of both the continuous and fluctuating AGN (although it is less visible for the latter).}.

The fact that the evolution of \N{OVI} converges in the limit of small \tagn, also means that we can calculate the offsets of \avNfossil{OVI} (and \avfcovfossil{OVI}) from equilibrium in the limit of small \tagn. We calculate these asymptotic offsets for $\fdutym = 1, 2, 5, 10, 20, 50 \%$ by considering continuously radiating AGN with $\eddratiom = 0.01, 0.02, 0.05, 0.1, 0.2, 0.5$, respectively, and show these values as separate columns in Fig.~\ref{fig:av_map_OVI} (called `Limit'). The asymptotic offsets generally increase with increasing \fduty, although they are slightly lower for $\fdutym = 50 \%$ than for $\fdutym = 20 \%$. This is due to the non-linear relation between \OVI enhancement and AGN luminosity (see also Section~\ref{sec:lum_dep}): the \OVI abundance in the newly established ionization balance between photoionization from the HM01 background and the AGN, collisional ionization and recombination, will \emph{increase} with increasing AGN luminosity if the AGN photoionizing flux is relatively low, as an increasing fraction of the oxygen is ionized from \OI{ }- \OV to \OVI. However, if the AGN flux is sufficiently high, the \OVI abundance will \emph{decrease} with increasing AGN luminosity, as an increasing fraction of the oxygen ends up in higher states than \OVI.

Returning to Fig.~\ref{fig:av_map_frac_OVI}, while for $t\sub{off} \lesssim 10$~Myr (to the left of the grey, dotted line) the \OVI is continuously kept in an overionized state, for $t\sub{off} \gtrsim 300$~Myr (to the right of the black, dashed line) there is hardly any AGN fossil effect at all. In this regime of long \tagn and small \fduty, the time in between two subsequent AGN-on phases is too long for an AGN-induced change in the \OVI to be observable for a significant fraction of the time.

Finally, in our post-processing calculation we neglect the effect of AGN photoheating, as we assume that the temperature of the gas is fixed. However, we expect this to have little impact on the CGM \OVI abundance, at least when the AGN is only on for a small fraction of the time. In Appendix~\ref{sec:photoheating}, we show that even in the limit where the time in between subsequent AGN-on phases is infinitely small -- adopting a continuously radiating AGN with $\eddratiom = 0.1$ in the \mstarsim{10} galaxy at $z = 3$ --, the effect of AGN photoheating on the \OVI abundance is negligibly small compared to the effect of AGN photoionization. We find that the latter is generally about an order of magnitude larger.


\section{Results for other metal ions}
\label{sec:other_metals}

\begin{figure*}
\begin{center}
\includegraphics[width=\textwidth]{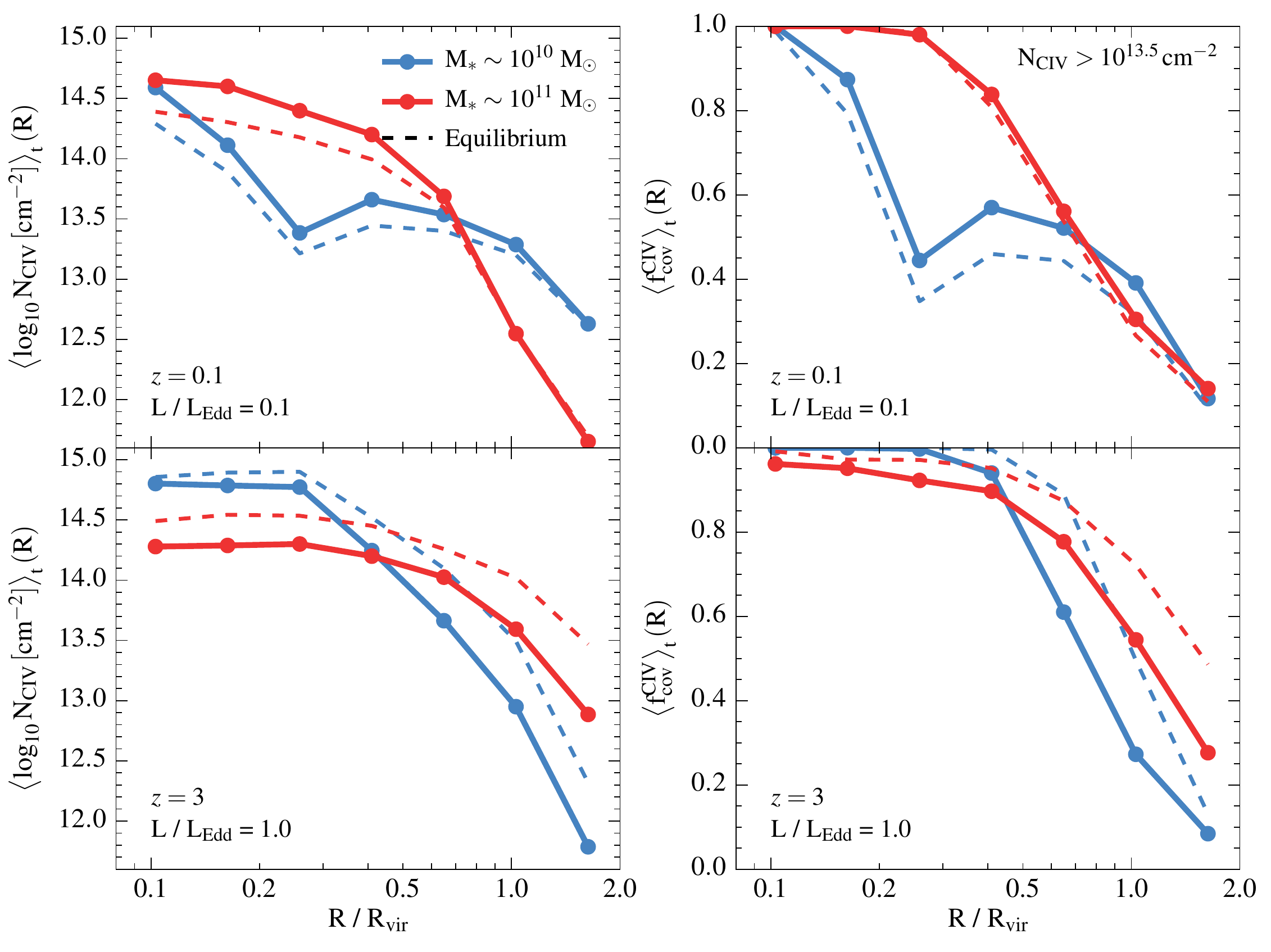}
\end{center}
\caption{As Fig.~\ref{fig:mass_dep_OVI}, but showing the fossil effect for \CIV. For the covering fraction, we adopt a column density threshold of $\Nm{CIV} > 10^{13.5}$ cm$^{-2}$. While \CIV is enhanced with respect to equilibrium at $z = 0.1$, \avN{CIV} and \avfcov{CIV} are lower than in equilibrium at $z = 3$. This is in contrast to \OVI, which is enhanced at both $z = 0.1$ and $z = 3$, and is due to the difference in ionization energy.}
\label{fig:mass_dep_CIV}
\end{figure*}

While this work focusses on \OVI, the abundances of other commonly observed ions -- like \SiIV and \CIV -- are also affected by fluctuating AGN. However, the strength of the fossil effect, and whether it induces an increase or decrease of the ion column density, depends on the ionization energies of the associated ions \citep{oppenheimer+schaye_2013b}. Since the ionization energies of \CIII and \SiIII are significantly lower than that of \OV, for the same AGN luminosity the flux of ionizing photons is higher. Hence, the AGN yields a higher ionization rate from the low carbon and silicon states to \CIV and \SiIV, respectively, but also from \CIV and \SiIV to higher states. This leads to \avN{CIV} and \avfcov{CIV} (adopting a column density threshold of $\Nm{CIV} > 10^{13.5}$ cm$^{-2}$) at $z=3$ that are, in contrast with \OVI, \emph{lower} than in equilibrium. However, at $z=0.1$, where the ionizing flux at a fixed $R / R\sub{vir}$ is lower, \avN{CIV} and \avfcov{CIV} are enhanced with respect to equilibrium. This is illustrated in Fig.~\ref{fig:mass_dep_CIV}, which shows a comparison between \avN{CIV}, \avfcov{CIV} and the respective profiles in equilibrium for the four galaxies, adopting the same AGN parameters as in Fig.~\ref{fig:mass_dep_OVI}. We note that the \CIV enhancement for the \mstarsim{11} galaxy at $z = 0.1$ is relatively small, in contrast to the large fossil effect seen for \OVI. This is due to the higher flux of \CIV-ionizing photons, which ionizes significant quantities of carbon to states higher than \CIV, thereby roughly canceling out the positive offset. Furthermore, while for our fiducial choice of $\tagnm = 10^6$~yr and $\fdutym = 10 \%$ the decrease in \avN{CIV} around the \mstarsim{10} galaxy at $z = 3$ is $\approx 0.1 - 0.5$~dex, for $\tagnm = 10^5$~yr and $\fdutym = 50 \%$ the decrease becomes as large as $\gtrsim 1.0$~dex.

For \SiIV (not shown), which has a $1.4$ times lower ionization energy than \CIV, the AGN causes a deficit of \avN{SiIV} and \avfcov{SiIV} with respect to equilibrium for all four galaxies, irrespective of the redshift. While the decrease in \avN{SiIV} around the \mstarsim{10} galaxy at $z = 3$ is $\approx 0.2 - 1.0$~dex for the fiducial $\tagnm = 10^6$~yr and $\fdutym = 10 \%$, the decrease is as large as $\approx 2.0 - 3.0$~dex for $\tagnm = 10^5$~yr and $\fdutym = 50 \%$.

For ions with ionization energies higher than \OVI -- like \NeVIII and \MgX -- the AGN fossil effect causes an enhancement of their column density and covering fraction, but the enhancement is less strong than for \OVI. This is due to the significantly lower flux of radiation above the ionization energy, which is required to enhance the ion column densities during the AGN-on phase. As a result, only for combinations of \tagn and \fduty for which $t\sub{off} \lesssim 1$~Myr ($t\sub{off} \lesssim 0.1$~Myr) we find \avN{NeVIII} (\avN{MgX}) to be offset from equilibrium by more than $0.1$~dex.


\section{Conclusions}
\label{sec:conclusions}

We have studied the impact of photoionization by fluctuating AGN on the \OVI abundance in the CGM of the host galaxies. We selected four galaxies from the \emph{Ref-L100N1504} simulation from the EAGLE project: two galaxies with stellar masses of \mstarsim{10} and \mstarsim{11} at $z = 0.1$, and two galaxies with similar stellar masses at $z = 3$. We implemented the sources of ionizing radiation in the centres of these galaxies in post-processing, and followed the time-variable abundances of $133$ ion species out to impact parameters of $2 R\sub{vir}$. Due to the significant recombination timescales of the oxygen ions, and the multiple levels that the ions need to recombine through, the central AGN leaves the CGM gas in an overionized state after it fades: this is what defines an AGN proximity zone fossil. This affects the CGM \OVI abundance, as one would for example measure in observations of quasar absorption-line systems, even though the galaxy would not be classified as an active AGN host at the moment of observation.

To quantify the significance of this fossil effect, we presented predictions of the average \OVI column density and covering fraction enhancement in between subsequent AGN-on phases, as well as the fraction of the time in between the two phases that the \OVI column density is out of equilibrium by at least $0.1$~dex. The latter quantity reflects the probability of observing a significant AGN fossil effect around a non-active AGN host. We investigated how these quantities depend on impact parameter, galaxy redshift, galaxy stellar mass, and on the AGN luminosity, lifetime and duty cycle fraction, where we explored a range of AGN parameters as constrained by observations. Our results can be summarized as follows.

\begin{enumerate}
\item For our fiducial choice of AGN luminosity ($\eddratiom = 0.1$ at $z = 0.1$ and $\eddratiom = 1.0$  at $z = 3$), lifetime ($\tagnm = 10^6$~yr) and duty cycle fraction ($\fdutym = 10 \%$), we find that all four galaxies are significantly affected by AGN fossil effects out to impact parameters of $R / R\sub{vir} = 2$. After the central AGN fades, the oxygen in the CGM is left in an overionized state for several megayears. For $R / R\sub{vir} < 1.3$, the next AGN-on phase starts before the gas can return to ionization equilibrium, keeping the \OVI abundance continuously enhanced. We find offsets in the time-averaged \OVI column density and covering fraction with respect to equilibrium that range from $\approx 0.3 - 1.0$~dex and $\approx 0.05 - 0.6$, respectively, at $R / R\sub{vir} \lesssim 0.3$ to $\approx 0.06 - 0.2$~dex and $\approx 0.05 - 0.1$ at $R / R\sub{vir} \approx 2$ (Fig.~\ref{fig:mass_dep_OVI}).
\item The AGN predominantly affects the photoionized gas at $T < 10^5$~K, while the change in the \OVI abundance in collionisally ionized gas ($T > 10^5$~K) is negligible. In the photoionized regime, the AGN increases the \OVI mass at $n\sub{H} = 10^{-4} - 10^{-1}$~cm$^{-3}$ at $z = 3$ ($n\sub{H} = 10^{-5} - 10^{-2}$~cm$^{-3}$ at $z = 0.1$). The affected density range is roughly constant as a function of impact parameter, even though the typical density of CGM decreases with increasing impact parameter. The re-equilibration timescale of \OVI, after the AGN turns off, is therefore also roughly independent of impact parameter (Fig.~\ref{fig:example_temp_dens_OVI}).
\item The dependence of the strength of the fossil effect on impact parameter, galaxy stellar mass and redshift follows from the difference in the ionizing photon flux, the abundance of \OI{ }- \OV oxygen ions and the \OVI re-equilibration timescale. The fossil effect is largest at small impact parameters for all four galaxies. At $z = 0.1$ and fixed $R / R\sub{vir}$, the fossil effect is stronger around the \mstarsim{11} galaxy than around the \mstarsim{10} galaxy, as a result of the $\approx 2$ times higher photon flux, the higher abundance of low-state oxygen ions and a similar re-equilibration timescale. At $z = 3$, the fossil effect is stronger around the lower-mass galaxy (except for $R / R\sub{vir} \gtrsim 1.3$), which is solely due to the higher abundance of low-state oxygen ions. While the gas at a fixed $R / R\sub{vir}$ at $z = 0.1$ receives a $60 - 80$ times lower photon flux than the gas around a galaxy with the same stellar mass at $z = 3$, the $\sim 10$ times longer re-equilibration timescale at $z = 0.1$ causes the \OVI column density and covering fraction to be strongly offset from equilibrium on average, yielding a particularly strong fossil effect for the \mstarsim{11} galaxy (Fig.~\ref{fig:mass_dep_OVI}).
\item The strength of the fossil effect tends to increase if the AGN lifetime is longer or if the duty cycle fraction larger, since the time in between subsequent AGN-on phases decreases. For the \mstarsim{10} galaxy at $z = 3$, \OVI is kept out of equilibrium continuously if $t\sub{off} \lesssim 10$~Myr. For these combinations of \tagn and \fduty, the \OVI enhancement accumulates over multiple cycles, until the gas eventually reaches a new (net) ionization equilibrium, where the combined photoionization from the HM01 background and the AGN (in addition to collisional ionization) balances the recombinations per cycle (Figs.~\ref{fig:av_map_OVI}, \ref{fig:av_map_frac_OVI} and \ref{fig:asymp_OVI}).
\item The strength of the fossil effect increases with increasing AGN luminosity. However, if the AGN luminosity becomes sufficiently high, the \OVI enhancement no longer scales linearly with AGN luminosity: for the \mstarsim{11} galaxy at $z = 0.1$, we find that the \OVI enhancement increases if we increase the Eddington ratio from $\eddratiom = 0.1$ to $\eddratiom = 1.0$, but not as significantly as if we increase the Eddington ratio from $\eddratiom = 0.01$ to $\eddratiom = 0.1$ (Fig.~\ref{fig:lum_dep_OVI}). This is due to the fact that the ionization state is sensitive to the ionization rates at all ions levels, hence to the rate from \OVI to \OVII as well as the rate from \OV to \OVI.
\item For low-ionization energy ions like \SiIV and \CIV, the AGN fossil effect causes a decrease in the CGM column density and covering fraction at $z = 3$, and an increase (\CIV) or a decrease (\SiIV) at $z = 0.1$ (Fig.~\ref{fig:mass_dep_CIV}). However, for the high-ionization energy ions \NeVIII and \MgX, we find only a significant fossil effect (enhancement) if $t\sub{off} \lesssim 1$~Myr and $t\sub{off} \lesssim 0.1$~Myr, respectively, which is due to the low flux of ionizing photons.
\item In the limit of short AGN lifetimes, the effect of a fluctuating AGN on the \OVI column density and covering fraction converges to the effect of a continuously radiating AGN with a luminosity equal to $(\fdutym / 100 \%)$ times the original luminosity (Fig.~\ref{fig:asymp_OVI}).
\end{enumerate}

Our results suggest that AGN proximity zone fossils are ubiquitous around $M\sub{\ast} \sim 10^{10 - 11} \, \mathrm{M}\sub{\odot}$ galaxies, and that these are expected to affect observations of metals in the CGM at both low and high redshifts. Since the AGN predominantly affect the low-temperature ($T < 10^5$~K), photoionized gas, fossil effects are expected to particularly alter the column densities of narrow absorption lines. Broad absorption lines, which mostly arise from high-temperature, collisionally ionized gas, will generally be insensitive to AGN fossil effects. However, since the total column density of an absorption system depends on the strength of both the photoionized and collisionally ionized components, the AGN affect measurements of the total column density, even if in ionization equilibrium most of the absorption is expected to be due to collisionally ionized gas. Furthermore, neglecting the impact of AGN fossil effects on the different metal ions may lead to significant errors in the inferred gas properties like density, metallicity and cloud size, when the observed absorption system is assumed to be in ionization equilibrium with the extra-galactic background.

We have shown that for our fiducial and observationally motivated choice of AGN parameters, the probability that a measurement of the CGM \OVI abundance is affected by AGN fossil effects, while the galaxy would not be identified as an active AGN host, is $\approx 100 \%$ out to impact parameters of at least one virial radius. The typical offsets in the \OVI column densities are comparable to the factor of $\approx 2 - 10$ discrepancy between the high \OVI columns found in observations and those predicted by simulations. This suggests that including the effect of fluctuating AGN in models of the CGM may be key to reproducing the observed abundances of metal ions. A detailed comparison, using a set of high-resolution EAGLE zoom simulations, with the \OVI observed around the $z \sim 0.2$ star-forming galaxies from the COS-Halos sample is presented by \citet{oppenheimer_2017}.


\section*{Acknowledgements}

This work used the DiRAC Data Centric system at Durham University, operated by the Institute for Computational Cosmology on behalf of the STFC DiRAC HPC Facility (www.dirac.ac.uk). This equipment was funded by BIS National E-infrastructure capital grant ST/K00042X/1, STFC capital grant ST/H008519/1, and STFC DiRAC Operations grant ST/K003267/1 and Durham University. DiRAC is part of the National E-Infrastructure. We also acknowledge PRACE for access to the resource Curie at Tr\'{e}s Grand Centre de Calcul. This work received financial support from the European Research Council under the European Union's Seventh Framework Programme (FP7/2007-2013) / ERC Grant agreement 278594-GasAroundGalaxies, from the UK STFC (grant numbers ST/F001166/1 and ST/I000976/1), and from the Belgian Science Policy Office ([AP P7/08 CHARM]). BDO's contribution was supported by NASA ATP grant number NNX16AB31G. AJR is supported by the Lindheimer Fellowship at Northwestern University.


\bibliographystyle{mnras}
\bibliography{bibliography}


\appendix

\section{Impact of AGN photoheating}
\label{sec:photoheating}

We compare the impact of photoheating by the radiation from the AGN on the CGM \OVI abundance to the effect of photoionization. The reaction network includes the option to explore photoheating effects from a time-variable source by allowing the gas temperatures to vary. In principle, we can explore these effects by comparing two network calculations, for the same set of AGN model parameters, with and without thermal evolution. However, one important complication is that the particle temperatures in EAGLE are not necessarily in thermal equilibrium to begin with. (Note that we work on simulation snapshot output(s) in post-processing.) They reflect the instantaneous state of the CGM, including shock-heated gas that is at a much higher temperature than expected based on its density. Once the temperature in the network calculation is allowed to vary, this gas will cool down, which will affect the abundance of \OVI, irrespective of whether an AGN is present.

\begin{figure}
\begin{center}
\includegraphics[width=\columnwidth]{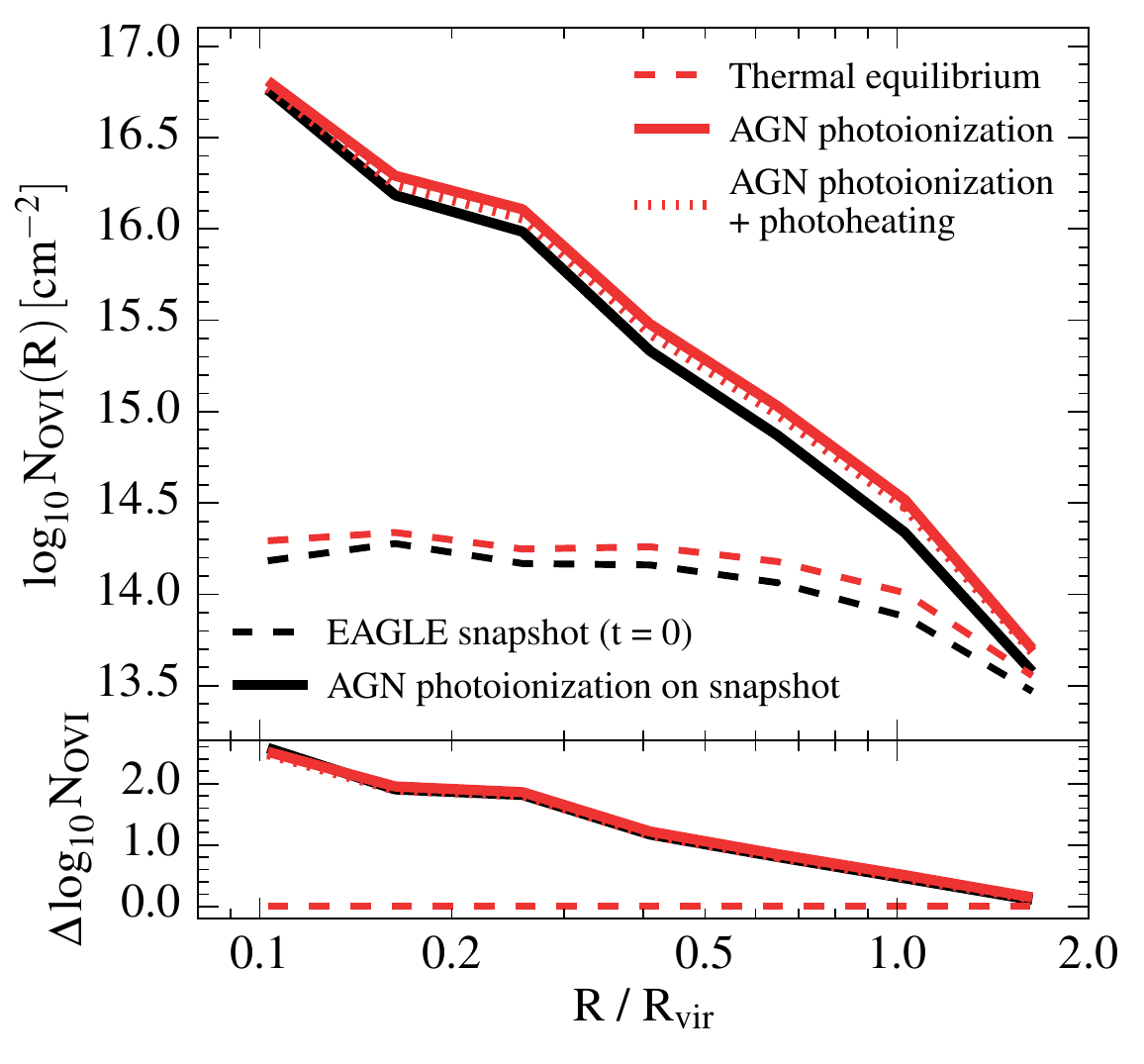}
\end{center}
\caption{The impact of AGN photoheating on the \OVI column density profile of the \mstar{1.0}{10} galaxy at $z = 3$, adopting a continuously radiating AGN with $\eddratiom = 0.1$. The red lines correspond to an artificial state with the $T < 10^5$~K gas in thermal and ionization equilibrium without AGN radiation (dashed) and in a new thermal and ionization equilibrium in the presence of the AGN radiation, where the gas temperature was kept fixed (solid) or allowed to vary (dotted). The temperature of the $T > 10^5$~K gas was kept fixed to the one in the EAGLE snapshot in both cases. For comparison, the black lines show the profile as calculated directly from the EAGLE snapshot assuming ionization equilibrium (dashed), and the profile corresponding to the new ionization equilibrium due to AGN photoionization using the snapshot temperatures (solid). The bottom panel indicates the difference in $\log\sub{10} \Nm{OVI}$ with respect to the dashed lines of the respective colour. The increase in \N{OVI} with respect to thermal equilibrium (red, dashed line) due to AGN photoheating and photoionization (red, dotted line) is only $\lesssim 0.05$~dex lower than the increase due to AGN photoionization (red, solid line), indicating that AGN photoheating does not have a significant impact on \N{OVI}.}
\label{fig:app_coldens}
\end{figure}

Hence, in order to isolate the effect of AGN photoheating (focusing on the \mstar{1.0}{10} galaxy at $z = 3$), we run a separate calculation, where we start from an artifical state with the $T < 10^5$~K gas in thermal (and ionization) equilibrium. For this gas, we run the reaction network including AGN radiation twice: once with the gas temperature kept fixed (hence, including only photoionization) and once where the temperature is allowed to vary (including both photoionization and photoheating). We consider a continuously radiating AGN with $\eddratiom = 0.1$ -- corresponding to a fluctuating AGN with $\eddratiom = 1.0$ and $\fdutym = 10 \%$ in the limit of small \tagn (see Section~\ref{sec:cycle_dep}) -- and let the gas evolve until it reaches a new ionization and thermal equilibrium. For the gas at $T > 10^5$~K, which is approximately the regime where gas must have been shock heated in the simulation, we fix its temperature to the one taken from the EAGLE snapshot, and only consider photoionization when AGN radiation is included. The red, dashed line in Fig.~\ref{fig:app_coldens} shows the \OVI column density profile for the initial state of this calculation (`Thermal equilibrium'). Since \OVI-bearing photoionized gas is generally very close to being in thermal equilibrium already, the column density profile is only slightly different from the one calculated directly from the EAGLE snapshot (assuming only ionization equilibrium; black, dashed line).

AGN photoionization increases \N{OVI} by $\approx 0.2 - 2.2$~dex (red, solid line). Including the effect of AGN photoheating slightly increases the temperature of the $T < 10^5$~K gas, by $\approx 0.23$~dex on average (weighted by the total oxygen mass) at $0.08 < R / R\sub{vir} < 2$. However, the resulting change in \N{OVI} is only modest: the increase with respect to thermal equilibrium due to AGN photoheating and photoionization (red, dotted line) is only $\lesssim 0.05$~dex lower than the increase due to just AGN photoionization. Hence, we conclude that AGN photoheating does not significantly affect the \OVI abundance in CGM gas, as the effect of AGN photoionization is generally larger by about an order of magnitude.


\bsp
\label{lastpage}
\end{document}